\documentclass[openacc]{rsproca_new}

\usepackage[english]{babel}

\usepackage{mathtools} 

\hypersetup{unicode=true}

\usepackage[sort]{natbib}
\usepackage[normalem]{ulem}

\usepackage[noprefix]{nomencl}
\usepackage{cancel}
\makenomenclature
\newcommand{\diff}[2]{\frac{\mathrm{d}{#1}}{\mathrm{d}{#2}}}
\newcommand{\pdiff}[2]{\frac{\partial{#1}}{\partial{#2}}}

\newcommand{\msz}[2]{\mathrlap{#1}\hphantom{#2}}
\newcommand{\hs}[1]{\hspace{#1mm}}
\newcommand{\etal}[0]{{\it et~al.~}}

\renewcommand{\vec}[1]{\boldsymbol{#1}}
\newcommand{\svec}[2]{\boldsymbol{#1}_{#2}}

\renewcommand{\d}{\;\mathrm{d}}
\newcommand{\intt}[2]{\int_{#1}^{#2}}

\newcommand{\e}[1]{\times 10^{#1}}

\newcommand{\mb}{\begin{pmatrix}}
\newcommand{\me}{\end{pmatrix}}

\definecolor{CustomBlue}{RGB}{0,38,166}
\definecolor{cB}{RGB}{0,60,200}
\definecolor{cR}{RGB}{250,0,0}
\definecolor{cG}{RGB}{38,200,0}
\definecolor{cO}{RGB}{240,150,0}
\definecolor{cLGy}{RGB}{240,240,240}

\definecolor{ref_col_1}{RGB}{0,80,0}
\definecolor{ref_col_2}{RGB}{120,0,0}

\newcommand{\SIN}{s_n}
\newcommand{\COS}{c_s}
\renewcommand{\st}{\SIN\theta}
\newcommand{\ct}{\msz{\,\COS}{\SIN}\theta}
\renewcommand{\ss}{\SIN\psi}
\newcommand{\cs}{\msz{\,\COS}{\SIN}\psi}
\renewcommand{\sp}{\SIN\phi}
\newcommand{\cp}{\msz{\,\COS}{\SIN}\phi}

\newcommand{\R}[2]{\vec{[R_{#1,#2}]}}
\newcommand{\methodName}{nonlinear beam shapes }
\newcommand{\MethodName}{Nonlinear beam shapes }
\DeclareMathSymbol{\Kappa}{\mathalpha}{operators}{"4B}
\newcommand{\zetaname}{kinematic vector }
\newcommand{\Scale}[2][4]{\scalebox{#1}{\ensuremath{#2}}}
\newcommand{\E}{\Scale[1.35]{\mathrm{e}}}

\newcommand*{\nolink}[1]{%
  {\textcolor{cB}{\protect\NoHyper#1\protect\endNoHyper}}%
}
\def\authorCntrb#1{{\vskip5.5pt\noindent \textcolor{jobcolor}{\fontsize{9}{11}\selectfont Author's contributions.}\fontsize{8}{11}\selectfont\enskip #1}}
\allowdisplaybreaks

\graphicspath{{./}}

\begin{document}

\title{On the Geometrically Exact Low Order Modelling of a Flexible Beam: Formulation and Numerical Tests}

\author{
C.~Howcroft$^1$, R.~G.~Cook$^1$, S.~A.~Neild$^1$, M.~H.~Lowenberg$^1$, J.~E.~Cooper$^1$, E.~B.~Coetzee$^2$}

\address{
$^1$ Dept. Aerospace Engineering, University of Bristol,\\
Queen's Building, University Walk, Bristol, BS8 1TR \\
$^2$ Future Projects Office, Airbus Operations Ltd,\\ Pegasus House,
Bristol, BS34 7PA}

\subject{nonlinear beam modelling, multibody dynamics}


\corres{C. Howcroft\\
\email{c.howcroft@bristol.ac.uk}}

\begin{abstract}
This paper proposes a low order geometrically exact flexible beam formulation based on the utilisation of generic beam shape functions to approximate distributed kinematic properties of the deformed structure. The proposed {\it \methodName} approach is in contrast to the majority of geometrically nonlinear treatments in the literature in which element based --- and hence high order --- discretisations are adopted. The kinematic quantities approximated specifically pertain to shear and extensional gradients as well as local orientation parameters based on an arbitrary set of globally referenced attitude parameters. In developing the dynamic equations of motion, an Euler angle parameterisation is selected as it is found to yield fast computational performance. The resulting dynamic formulation is closed using an example shape function set satisfying the single generic kinematic constraint. The formulation is demonstrated via its application to the modelling of a series of static and dynamic test cases of both simple and non-prismatic structures; the simulated results are verified using MSC Nastran and an element-based intrinsic beam formulation. Through these examples it is shown that the \methodName approach is able to accurately capture the beam behaviour with a very minimal number of system states.
\end{abstract}
\maketitle
\section{Introduction}
Within the field of flexible structural modelling, one dimensional beam models display a surprisingly wide applicability to the representation of numerous structural applications such as the modelling of aircraft wings, flexible satellites, turbine blades, communication towers, energy harvesters, atomic force microscopes and DNA molecules, to name but a few. Such treatments are typically suited to relatively slender structures with a single characteristic large dimension and, together with an appropriate constitutive material law, a one-dimensional representation may capture many aspects of the full three-dimensional problem. The same slender properties that make many physical structures amenable to beam representation often result in significant structural flexibility and, particularly for systems with free boundary conditions, this can lead to large deflections and subsequent geometric nonlinearity. Therefore, many studies treat such nonlinearity in kinematic frameworks permitting arbitrarily large rotation of either the structure directly, or within element frames in which structural sub-domains are treated.
Of particular interest to the present study is the class of geometrically exact modelling methods first proposed in the seminal work of Reissner \cite{Reissner1973}. These formulations are centred around the concept of a local body fixed coordinate system in which the material deformation of the beam is cast. The kinematic description of the beam follows from knowledge of this local material deformation, as well as the span-varying orientation of the local body system, which may comprise a large rotation from the initial undeformed configuration; consequently a global formulation of the flexible beam is attained in which geometric nonlinearity is inherently captured.

In a majority of cases, such geometrically nonlinear formulations are implemented using an element based discretisation of the structure. In \cite{Simo_Pt1,Simo_Pt2}, Simo and Vu-Quoc present a geometrically exact parameterisation based upon the vector components of the global rotation and displacement from the initial undeformed configuration. This approach yields a geometrically nonlinear finite-element formulation of the flexible structure in terms of rotation and displacement states. In \cite{Bauchau2014} Bauchau \etal compare this formulation with the absolute nodal coordinate formulation of Shabana \etal \cite{Shabana1998} that utilises gradient information rather than a rotational field in expressing the deflected geometry of the system. With a sufficiently fine discretisation, linear strain-displacement laws may be utilised within each element frame. The class of co-rotational formalisms (first proposed in \cite{Wempner1969,Belytschko1973,Belytschko1979}) take advantage of this by combining small local material deformation with arbitrarily large total rotations of the structure via the treatment of an element fixed (co-rotational) reference frame. In \cite{Cesnik2002}, Cesnik and Brown allow for larger local deformations by accounting for the non-small linear variation of curvature within each element frame; the resulting formulation is cast directly in terms of curvature rather than displacement states. Consequently the larger deformation allowed within each element frame affords the use of fewer elements in capturing geometric nonlinearity at the cost of increased complexity and bandwidth of the system matrices.
In \cite{Santos2011_etal} Santos \etal employ stress-resultant and displacement states in formulating a complimentary energy-based method for the representation of large static deformations. Together with additional displacement constraints this allows for the treatment of nonlinearly deformed frame structures, free from the shear-locking phenomenon \cite{Prathap1982} and allowing for the accurate recovery of distributed stress. However, this comes at the cost of 24 degrees of freedom per element (see \cite{Santos2011_overview} for an overview of past research efforts and ongoing challenges in developing complementary energy principles for large deformation problems).
In \cite{Hodges1990}, Hodges introduces the intrinsic beam formulation wherein constituent relations and the subsequent equations of motion are cast purely in a span-varying body referenced intrinsic coordinate system. In \cite{Patil2001}, Patil \etal utilise an element-based implementation of this formulation in studying the static aeroelastic response of a simple high aspect ratio aircraft. Here, the intrinsic reference system effectively becomes the co-rotational system for each element and global geometric information is recovered via quaternion integration from root to tip. The review papers of Wasfy and Noor \cite{Wasfy2003} and Shabana \cite{Shabana1997} further capture the large body of flexible multibody dynamic modelling literature.

A key feature of the above element-based descriptions is the generality with which structural configurations may be represented; however, this comes at the expense of a large number of system states. To reduce the problem size, one approach is to use model reduction methods; here one projects the full system onto a subset of modes using a regression analysis (for example via implicit condensation \cite{Hollkamp2008,Kuether2016} or enforced displacement \cite{Przekop2012} tests; see the review of Mignolet \etal \cite{Mignolet2013} for further details) to fit the coefficients of higher order terms, thus capturing nonlinear effects. However, the degree of nonlinear truncation required to render the regression problem tractable, as well as the inherent errors introduced when approximating the transient dynamics of the system, limits the domain of validity over which reduction techniques suitably approximate the full system.

For the current study a different approach is adopted. Rather than reducing a high order (large number of states) element-based model, the focus is cast instead on the low order (minimal state) approximation of the full geometrically nonlinear dynamic problem. To this end a {\it \methodName} approach is employed which approximates the kinematic quantities distributed along the full length of the flexible structure without subdivision into smaller elements. Compared to a typical element-based representation the reduction in system states can be significant. Such an approach is typical for the treatment of linear systems via for example Rayleigh Ritz, Galerkin and Modal type approximations (see standard texts on vibration modelling e.g. \cite{Rao2007}). In a geometrically nonlinear context, Patil and Althoff \cite{Patil2010} consider higher order shape functions as applied to the intrinsic beam formulation of \cite{Hodges1990}; here a set of shifted Legendre functions are used in approximating local angular velocity, translational velocity, force and moment quantities, expressed in the intrinsic reference frame. The resulting formulation is applied to the modelling of a flexible beam; however, a number of simplifying assumptions are required (i.e. treatment of a prismatic, constant property beam) in rendering the problem tractable to analysis. In the current paper, a geometrically exact flexible beam formulation is presented using a shape-based approximation of the kinematic beam quantities to yield a low order continuous formulation of the problem. By using a global referencing attitude representation of the beam orientation at any spanwise location, analyticity is retained up to and including the integrands of positional quantities and their derivatives, thus lessening the computational requirements on the numerical integration of terms. The resulting dynamic equations are suitable for the representation of beam problems displaying generic span varying geometric and material properties that typically exist in real structures; the underlying shape functions are kept general with candidate sets requiring only the satisfaction of a single kinematic constraint.

This paper is divided into two parts detailing firstly the modelling approach, followed by examples of its application. In section \ref{sec: Presentation of Dynamic Equations} the mathematical formulation of the dynamic equations is presented, beginning with the variational statement of the geometrically exact problem and then introducing the chosen parameterisation. The generic requirements and example application of an approximating shape basis is detailed in section \ref{sec: Shape Function Basis}. In section \ref{sec: Numerical Test Cases} a variety of validating test cases, which exercise various aspects integral to the treatment of a typical flexible beam problem, are presented. Throughout these test cases, additional results are generated as a baseline for comparison using both MSC Nastran (a well established finite-element package capable of treating geometric nonlinearity) and the intrinsic beam formulation of \cite{Hodges1990} implemented as a finite-element code (see \cite{Hodges1996,Palacios2010}). The latter modelling approach forms a particularly suitable base of comparison as, in addition to constituting a well documented and validated nonlinear treatment, the resulting equations of motion take the form of a high order system with fairly sparse system matrices and maximum order 2 nonlinearity. This serves as an interesting counterpart to the proposed formulation in this study which admits a greatly reduced number of states with fully populated form and a high degree of nonlinearity. Following the demonstration of the proposed \methodName approach in treating the discussed examples, conclusions are drawn in section \ref{sec: Conclusions}.

\section{Presentation of Dynamic Equations}\label{sec: Presentation of Dynamic Equations}
This section presents the derivation of the equations of motion describing a generic flexible nonlinear beam. This derivation proceeds with a geometrically exact beam treatment \cite{Simo_Pt1,Simo_Pt2,Antman1974} wherein consideration of a local intrinsic coordinate system is used to capture the full geometric nonlinearity inherent in this class of flexible beam problem.
By relating generic shape function sets to a spanwise varying \zetaname --- consisting of three attitude, one extension and two shear parameters --- a minimal state representation of the system is developed here which yields fast computational performance, easy truncation of shear and extensional deformations and admits representation as a nonlinear system of ordinary differential equations.

In sections \nolink{\ref{sec: Presentation of Dynamic Equations}}\ref{sec: Kinematic Description} and \nolink{\ref{sec: Presentation of Dynamic Equations}}\ref{sec: Virtual Work Terms} a general statement of the geometrically exact variational problem is given wherein virtual work contributions are presented in terms of incremental displacement, shear, extensional and curvature deformations of the beam.
In section \nolink{\ref{sec: Presentation of Dynamic Equations}}\ref{sec: Attitude Representation} an Euler angle attitude representation is used to link these variational terms to a minimal set of shear, extension and rotational parameters. Following this, in section \nolink{\ref{sec: Presentation of Dynamic Equations}}\ref{sec: Generalised Coordinates}, a shape function discretisation is used to split these parameters into spacial and temporal components yielding a minimal state representation of the problem in terms of the set $\vec{q}$ of time dependent generalised states from which the equations of motion follow.

\subsection{Kinematic Description}\label{sec: Kinematic Description}

\begin{figure}[h]
\centering
\def\svgwidth{0.9\linewidth}
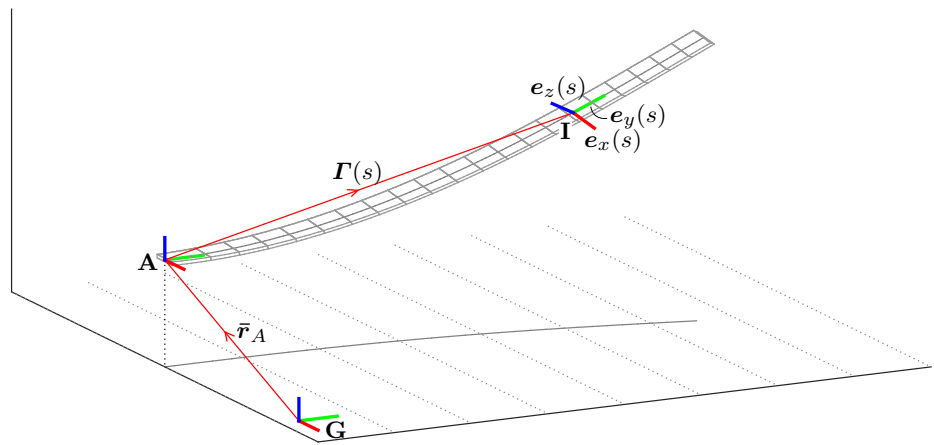
\caption{Reference coordinate systems and position vectors used in characterising the 3D beam geometry.}
\label{fig: geometric_definitions}
\end{figure}

\begin{figure}[t]
\centering
\def\svgwidth{0.9\linewidth}
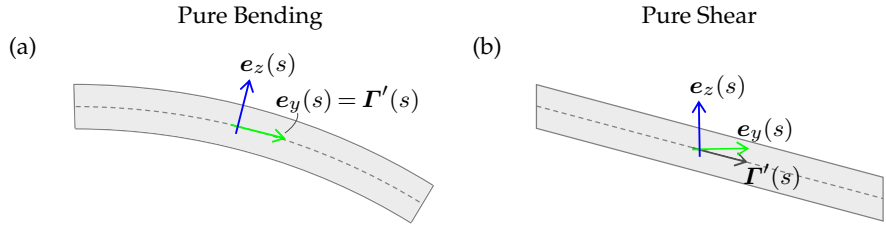
\caption{Illustration of the distinction between bending and shear type structural deformation.}
\label{fig: bending_vs_shear}
\end{figure}

Consider first the flexible beam of length $L$ depicted in figure \ref{fig: geometric_definitions}, deemed to be a suitable representation of some slender three-dimensional structure, reduced about an arbitrary material reference line that forms the beam axis; the curvilinear coordinate $s\in[0,L]$ is used to parameterise the distance along this axis. The axis itself is defined by the position vector $\vec{\Gamma}(s)$ taken in some arbitrary reference system denoted $\bf A$ (here considered with origin placed at the beam root). This system is in turn embedded in the inertial reference system $\bf G$ with positional offset $\svec{\bar r}{A}$. Now take the intrinsic body fixed system $\bf I$ (depicted at an example location $s$ along the beam) represented by the orthonormal vector triad $\{\svec{e}{x},\svec{e}{y},\svec{e}{z}\}$. In the undeformed configuration, $\svec{e}{y}$ lies parallel to the beam axis and $\svec{e}{x}$ and $\svec{e}{z}$ are mutually perpendicular directions in the cross sectional plane normal to $\svec{e}{y}$. As the beam deflects the orientation of $\svec{e}{x}$ and $\svec{e}{z}$ (in the absence of cross section warping) follow the evolution of their respective material lines whilst $\svec{e}{y}$ remains normal to the cross section plane they span. In figure \ref{fig: bending_vs_shear} example orientations of $\svec{e}{y}$, $\svec{e}{z}$ and $\vec{\Gamma}^\prime(s)$ (where the superscript $\bullet^\prime$ denotes differentiation with respect to $s$) are shown for the cases of planar pure bending (a) and pure shear (b). The difference in cross section orientation for each sub-case illustrates the important point that, in the presence of shear deformation, $\svec{e}{y} \neq \vec{\Gamma^\prime}$. Therefore in general the reference line $\vec{\Gamma}^\prime(s)$ comprises a spanwise integration along $\svec{e}{y}$ (including axial extension) in addition to shear deformations in the $\svec{e}{x}$ and $\svec{e}{z}$ directions. $\svec{\Gamma}{[G]}(s)$ is thus given the form:
\begin{equation}
    \svec{\Gamma}{[G]}(s) = \svec{\bar r}{A[G]} + \R{G}{A}\intt{0}{s} (1+\varepsilon(\tilde s))\,\svec{e}{y[A]}(\tilde s) + \tau_x(\tilde s)\,\svec{e}{x[A]}(\tilde s) + \tau_z(\tilde s)\,\svec{e}{z[A]}(\tilde s) \d \tilde s
    \label{eq:gamma}
\end{equation}
(compare with Reissner \cite{Reissner1973}) where $\varepsilon(s)$, $\tau_x(s)$ and $\tau_z(s)$ denote axial extension, $\svec{e}{x}$ shear and $\svec{e}{z}$ shear deformations of the beam at each spanwise location, respectively.
Each subscript in square brackets denotes the reference system ($\bf G$, $\bf A$ or $\bf I$ depicted in figure \ref{fig: geometric_definitions}) in which each vector quantity is cast. These coordinate systems are related by the rotation matrices $\R{G}{A}$ and $\R{A}{I}$ such that for any vector $\vec{v}$
\begin{equation}
    \svec{v}{[G]} = \R{G}{A}\svec{v}{[A]} = \R{G}{A}\R{A}{I}\svec{v}{[I]}\;.
\end{equation}

\subsection{Virtual Work Terms}\label{sec: Virtual Work Terms}

Following on from the kinematic description, the equations of motion describing the geometrically exact beam are now developed in weak form; individual terms are formulated via the principle of virtual work done by internal and external forces acting over incremental displacements, rotations and beam strains. These work terms are detailed in the following subsections and in each case, the variational quantities are related back to the intrinsic coordinate system depicted in figure \ref{fig: geometric_definitions}.

\subsubsection{Material Stress}

In treating the stress terms of the system, strain deformations of the structural material are considered as lumped translational and rotational gradients along the parametric beam line $\vec\Gamma(s)$. Hence the deformation vector $\vec\xi(s)$ is introduced here which consists of the components
\begin{equation}
    \vec{\xi}(s) = \left( \;\tau_x(s) \,,\, \varepsilon(s) \,,\, \tau_z(s) \,,\, \kappa_x(s) \,,\, \kappa_y(s) \,,\, \kappa_z(s)\;\right)^T\;.
    \label{eq:xi_flat}
\end{equation}

Here, $\kappa_x(s)$, $\kappa_y(s)$ and $\kappa_z(s)$ give the $\svec{e}{x}$, $\svec{e}{y}$ and $\svec{e}{z}$ components of the beam curvature $\vec\Gamma^{\prime\prime}(s)$, respectively. The curvature may be written\footnote{Note that this curvature does not depend on the axial extension $\varepsilon$ and therefore represents the rotational gradient per undeformed arclength $s$.}
\begin{equation}
  \vec{\kappa} = \left(\kappa_x\,,\,\kappa_y\,,\,\kappa_z\right)^T =
  \left(\svec{e^\prime}{y}\cdot\svec{e}{z}\,,\,
         \svec{e^\prime}{z}\cdot\svec{e}{x}\,,\,
   \svec{e^\prime}{x}\cdot\svec{e}{y}\right)^T
   \label{eq:kappa_E}
\end{equation}
often expressed in the compact form
\begin{equation}
    \vec{\tilde\kappa} = \vec{\E^T}\vec{\E^\prime}\hs{6},\hs{14}
    \vec{\E} = \mb\svec{e}{x}\mid\svec{e}{y}\mid\svec{e}{z}\me
\end{equation}
where $\vec{\tilde\kappa}$ denotes the skew symmetric matrix formed from the vector $\vec\kappa = \mb\kappa_x\,,\,\kappa_y\,,\,\kappa_z\me$.

With these curvatures defined, the strain-induced internal force and moment distribution may be written
\begin{equation}
  \svec{F}{K} = -\vec{\tilde K}\vec{\Delta\xi}\hs{6},\hs{14}
  \vec{\Delta\xi} = (\vec{\xi}-\vec{\xi_0})
\end{equation}
where $\vec{\xi_0}$ denotes some minimal stress pre-curvature of the beam and the generic stiffness matrix is written
\begin{flalign}
    \vec{\tilde K} = \mb
    K_{11}(s,\vec{\Delta\xi}) & K_{12}(s,\vec{\Delta\xi}) & \cdots & K_{16}(s,\vec{\Delta\xi})\\
    K_{21}(s,\vec{\Delta\xi}) & K_{22}(s,\vec{\Delta\xi}) & \cdots & K_{26}(s,\vec{\Delta\xi})\\
    \vdots                    & \vdots                    & \ddots & \vdots                   \\
    K_{61}(s,\vec{\Delta\xi}) & K_{62}(s,\vec{\Delta\xi}) & \cdots & K_{66}(s,\vec{\Delta\xi})\\ \me
    \qquad K_{ij} = K_{j\,i}\;\;\;\;.
\end{flalign}
For the examples of this study the linear stress-strain relationship is assumed wherein $\mathrm{diag}(\vec{\tilde K}) = (GA,EA,GA,EI_{xx},GJ,EI_{zz})$ and all other entries are zero. Note however that, more generally, one may simply consider
\begin{equation}
  \svec{F}{K} = \svec{F}{K}(s,\vec{\Delta\xi})\;,
  \label{eq: F_K_general}
\end{equation}
to be some arbitrary constitutive law relating the internal stress distribution to deformations $\vec{\Delta\xi}$ about the beam reference axis $\vec\Gamma(s)$ \cite{Antman1974}. For structures with simple geometries and known material properties this may be obtained analytically by relating the beam deformation vector $\vec\xi$ to the material strain tensor (see for example \cite{Hodges1974}); for more complex structures the constitutive relation may be obtained from physical or modelled tests of the full structure followed by the subsequent one-dimensional reduction about an arbitrary reference line \cite{Yu2002}.
In this general case of equation (\ref{eq: F_K_general}) the virtual work done over the beam, given the incremental strain deformation $\vec{\delta\xi}$, is
\begin{equation}
    \delta W_K = \intt{0}{L}\svec{F}{K}(s,\vec{\Delta\xi}) \cdot \vec{\delta\xi} \d s\;.
    \label{eq:delta W_K}
\end{equation}

\subsubsection{Structural damping}
Formulation of the virtual work done by structural dissipation follows in much the same form as the preceding strain term. Considering again the variation $\vec{\delta\xi}$, one may specify the function $\svec{F}{C}(\vec{\Delta\xi},\vec{\dot\xi)}$ relating the internal structural damping force $\svec{F}{C}$ to the strain rates $\vec{\dot\xi}$. Thus the virtual work for this incremental deformation takes the form
\begin{equation}
    \delta W_C = \intt{0}{L}\svec{F}{C}(s,\vec{\Delta\xi},\vec{\dot\xi}) \cdot \vec{\delta\xi} \d s
    \;,\hs{10}
    \vec{\dot{\tilde\kappa}} = \vec{\dot \E^T}\vec{\E^\prime} + \vec{\E^T}\vec{\dot \E^\prime}
    \label{eq:delta W_C}\;.
\end{equation}

\subsubsection{Applied Loads}
{\bf External Forces:}
The work done by an applied span varying force vector $\svec{F}{[G]}(s)$ with components in the global reference system is expressed with respect to the infinitesimal displacement of the reference line $\svec{r}{F[G]}$ along which it is applied. This reference line in the global system is given by
\begin{flalign}
    \svec{r}{F[G]}(s) &= \svec{\bar r}{A[G]} + \R{G}{A}\svec{\Gamma}{[A]}(s) + a_F(s)\svec{e}{x[G]}(s) + c_F(s)\svec{e}{z[G]}(s)\;.
    \label{eq:r_F}
\end{flalign}
Recall $\svec{\bar r}{A[G]}$ from figure \ref{fig: geometric_definitions} and define $a_F$ and $c_F$ as the offset of $\svec{r}{F[G]}$ from the beam reference line $\vec\Gamma$ in the $\svec{e}{x}$ and $\svec{e}{z}$ directions, respectively. This reference line follows the deformation of the beam structure and, in the absence of cross-sectional warping, $a_F$ and $c_F$ remain independent of the beam strain.
Thus, the virtual work due to the applied load $\svec{F}{[G]}(s)$ takes the form
\begin{equation}
    \delta W_F = \intt{0}{L} \svec{F}{[G]}(s) \cdot \svec{\delta r}{F[G]}(s) \d s\;,
    \label{eq:delta W_F}
\end{equation}
where the variational change in reference line is
\begin{flalign}
    \svec{\delta r}{F[G]}(s) = \svec{\delta\bar r}{A[G]}
     + \svec{\delta \Gamma}{[G]}(s)
     + a_F(s)\vec{\delta}\svec{e}{x[G]}(s) + c_F(s)\vec{\delta}\svec{e}{z[G]}(s)
    \label{eq:delta r_F}
\end{flalign}
and
\begin{flalign}
    \svec{\delta \Gamma}{[G]}(s) = 
    \intt{0}{s}
    &\left((1+\varepsilon(\tilde s))\,\vec{\delta}\svec{e}{y[G]}(\tilde s) + \delta\varepsilon(\tilde s)\svec{e}{y[G]}(\tilde s)
  + \tau_x(\tilde s)\,\vec{\delta}\svec{e}{x[G]}(\tilde s)\right. \notag\\
  &+ \left.\delta\tau_x(\tilde s)\,\svec{e}{x[G]}(\tilde s)
  + \tau_z(\tilde s)\,\vec{\delta}\svec{e}{z[G]}(\tilde s) + \delta\tau_z(\tilde s)\,\svec{e}{z[G]}(\tilde s)\right) \d \tilde s\;.
  \label{eq:delta gamma}
\end{flalign}
Note that $F$ is general, in that its magnitude and orientation may vary arbitrarily in time and/or depend upon the states of the system, e.g. as in the case of a time-varying follower force.
\\ \\
{\bf External Moment:}
Applied moments are denoted by the distributed vector $\vec{M}(s)$ for which the work done is considered with respect to the rotational variation $\vec{\delta\vartheta}(s)$. This variation is written as
\begin{equation}
    \svec{\delta\vartheta}{[G]}(s) = \frac{1}{2}(\svec{e}{x}(s)\times\vec\delta\svec{e}{x}(s) + \svec{e}{y}(s)\times\vec\delta\svec{e}{y}(s) + \svec{e}{z}(s)\times\vec\delta\svec{e}{z}(s))_{[G]}
    \label{eq:vartheta_ecross}
\end{equation}
and thus
\begin{equation}
    \delta W_{M} = \intt{0}{L} \svec{M}{[G]}(s) \cdot \svec{\delta\vartheta}{[G]}(s) \d s\;.
    \label{eq:delta W_M}
\end{equation}

\subsubsection{Kinetic terms}
Consider the reference line $\svec{r}{m}(s)$ intersecting the centre of mass of all infinitesimal cross sections along the length of the beam.
The inertial terms are furnished by considering the work performed when imparting a change in the momentum of the beam over incremental displacements $\svec{\delta r}{m}(s)$ and rotations $\vec{\delta\vartheta}(s)$ of this reference line. The virtual work is thus written:
\begin{flalign}
  \delta W_T &= -\intt{0}{L} \diff{}{t}\left(
    m(s)\svec{\dot r}{m[G]}(s)\right)\cdot\svec{\delta r}{m[G]}(s) \d s \notag\\
             &\msz{}{=}- \intt{0}{L} \diff{}{t}\left(
    \svec{\E}{[G]}(s)\svec{I}{\vartheta[I]}(s)\vec{\E_{[G]}^T}(s)\svec{\dot\vartheta}{[G]}(s)\right)\cdot\svec{\delta\vartheta}{[G]}(s) \d s \notag \\[2mm]
             &= \intt{0}{L} -\left(m\svec{\ddot r}{m[G]}\cdot\svec{\delta r}{m[G]}
             + \svec{\dot\E}{[G]}\svec{I}{\vartheta[I]}\vec{\E_{[G]}^T}\svec{\dot\vartheta}{[G]}\cdot\svec{\delta\vartheta}{[G]}\right. \notag\\[-2mm] &\msz{}{=\intt{0}{L}}+
    \cancelto{0}{\svec{\E}{[G]}\svec{I}{\vartheta[I]}\vec{\dot\E_{[G]}^T}\svec{\dot\vartheta}{[G]}}\cdot\svec{\delta\vartheta}{[G]} +
    \left.\svec{\E}{[G]}\svec{I}{\vartheta[I]}\vec{\E_{[G]}^T}\svec{\ddot\vartheta}{[G]}\cdot\svec{\delta\vartheta}{[G]}\right) \;\d s\;,
             \label{eq:delta W_T}
\end{flalign}
where the first and second terms correlate to the linear and angular momentum, respectively. Expressions for $\svec{r}{m}(s)$ and its variation $\delta\svec{r}{m}(s)$ take the identical form to equations (\ref{eq:r_F}) and (\ref{eq:delta r_F}) with the substitution $\{a_F,c_F\}\rightarrow\{a_m,c_m\}$.
$m(s)$ gives the mass per unit length along $\svec{r}{m}(s)$ and $\svec{I}{\vartheta}(s)$ the sectional inertia matrix. Additional point masses may be incorporated via substitution of the appropriate Dirac delta representations onto the reference line $\svec{r}{m}(s)$ and mass distribution $m(s)$.
The remaining accelerations $\vec{\ddot\vartheta}$ and $\svec{\ddot r}{m}$ are detailed in the appendices.

\subsection{Attitude Representation}\label{sec: Attitude Representation}
To further develop the describing equations of motion, one must relate the variational quantities $\vec{\delta\Gamma}$, $\vec{\delta\xi}$ and $\vec{\delta\vartheta}$ to a consistent set of kinematic parameters over which the problem will be solved. Together these variations depend upon the deformation of the flexible beam as well as the translational and rotational motion of the reference system ${\bf A}$.

For the deformation of the flexible structure one may consider the components of the vector $\vec\xi = ( \;\tau_x(s) \,,\, \varepsilon(s) \,,\, \tau_z(s) \,,\, \kappa_x(s) \,,\, \kappa_y(s) \,,\, \kappa_z(s)\;)^T$ in the material frame as a natural description of this deformation. However, although the above variations are uniquely determined by these components, there is no closed form solution directly relating the curvature in the material frame to $\svec{\vartheta}{[A]}$ and $\svec{\Gamma}{[A]}$ for all but the simplest $\kappa_x(s)$, $\kappa_y(s)$, $\kappa_z(s)$ distributions (note that Cesnik \etal circumvent this issue by assuming simplified $\kappa$ approximations within individual element frames, building up complexity via discretisation \cite{Cesnik2002}).

\begin{figure}[t]
\centering
\def\svgwidth{0.7\linewidth}
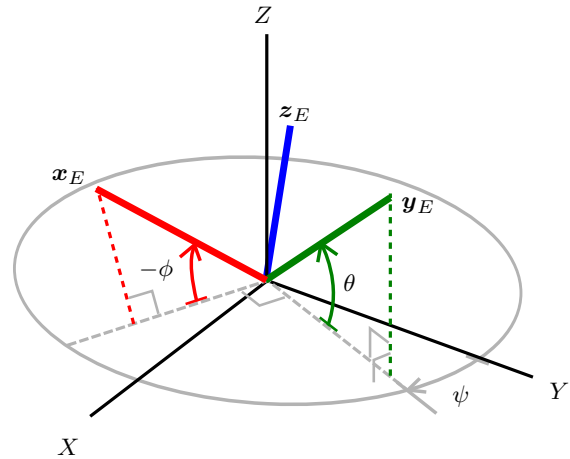
\vspace{-0.6cm}
\caption{The set of Euler angles used to parameterise the intrinsic system orientation.}
\label{fig: euler_angles}
\end{figure}

For the continuous formulation developed here, this lack of a closed form relation complicates the treatment of these quantities and their derivatives. To address this issue one notes from equations (\ref{eq:gamma}), (\ref{eq:vartheta_ecross}), (\ref{eq:xi_flat}) and (\ref{eq:kappa_E}) that these variations (in the reference system ${\bf A}$) are dependent on the intrinsic coordinate system $\svec{\E}{[A]}$ such that:
\begin{flalign}
    \svec{\Gamma}{[A]}(s) &= \svec{\Gamma}{[A]}\left(\,\tau_x(s) \,,\, \varepsilon(s) \,,\, \tau_z(s) \,,\, \svec{\E}{[A]}(\vec p)\,\right)\;,\notag\\
    \svec{\vartheta}{[A]}(s) &= \svec{\vartheta}{[A]}\left(\,\svec{\E}{[A]}(\vec p)\,\right)\;,\notag\\
    \svec{\xi}{[A]}(s) &= \svec{\xi}{[A]}\left(\,\tau_x(s) \,,\, \varepsilon(s) \,,\, \tau_z(s) \,,\, \svec{\E}{[A]}(\vec p)\,,\, \svec{\E^\prime}{[A]}(\vec p) \,\right)\;,
\end{flalign}
where $\vec{p}$ denotes the set of attitude parameters uniquely defining $\svec{\E}{[A]}$. There are numerous potential choices for such a parameterisation including, for example, quaternions, rotation vector, Rodrigues parameters; these and other such parameterisations are detailed in \cite{Shuster1993}. For the remainder of this study an example set of asymmetric 3-1-2 Euler angles are employed; like other three-parameter descriptions this has the advantage of a minimal state representation and in the context of the current formulation is found to yield fast computational performance. The attitude states --- here denoted $\vec{p} = (\theta, \psi, \phi)$ --- relating to this parameterisation are depicted in figure \ref{fig: euler_angles}. The indicated angular deflections are labelled in a consistent manner with the particular positive convention employed in equation (\ref{eq:E[A]}).
Given these angular definitions, and taking the reference system ${\bf A}$ such that its $X$,$Y$ and $Z$ axis vectors coincide with the beam root vectors $\svec{e}{x}(0)$, $\svec{e}{y}(0)$ and $\svec{e}{z}(0)$ respectively, one may write
\begin{flalign}
  \svec{\E}{[A]} = &\mb\svec{e}{x[A]} \mid \svec{e}{y[A]} \mid \svec{e}{z[A]}\me
  = \svec{R}{E}\mb\svec{x}{E} \mid \svec{y}{E} \mid \svec{z}{E}\me\vec{R_E^T} \notag \\
  =
  \svec{R}{E}
  \hs{-1}&\mb
    \msz{}{-}\cs\cp+\st\ss\sp & \ct\ss & \msz{}{-}\cs\sp-\st\ss\cp \\
    -\ss\cp+\st\cs\sp & \ct\cs & -\ss\sp-\st\cs\cp \\
    -\ct\sp & \st & \ct\cp
  \me\hs{-1.2} \vec{R_E^T}
  \label{eq:E[A]}
\end{flalign}
\\
where the shorthand convention $\SIN$ and $\COS$ for the trigonometric operations `$\sin()$' and `$\cos()$' is used; $\svec{R}{E}$ denotes the optional transform specifying the mapping from the Euler axis $(\svec{x}{E},\svec{y}{E},\svec{z}{E})$ to the intrinsic system $(\svec{e}{x},\svec{e}{y},\svec{e}{z})$. Via the choice of $\svec{R}{E}$ one may place the $\theta = \pm 90\deg$ Euler singularities of the problem at generic opposing orientations of the intrinsic reference frame ${\bf A}$. For example, when $\svec{R}{E}$ takes the value of the 3$\times$3 identity matrix, the Euler singularities fall at intrinsic frame orientations satisfying the equality $\svec{e}{y[A]} = (0,\pm1,0)^T$.

Note that an additional attitude and position parameterisation can be assigned to the motion of system $A$ with respect to the global reference ${\bf G}$ for problems where the root kinematics are not prescribed. This is useful in treating free problems e.g. the modelling of a flexible aircraft in free flight \cite{Cesnik2002,Patil2001,Palacios2010,Hesse2013}. For brevity these terms are not furnished in this derivation but are easily appended to the problem; therefore from here $\svec{\bar{r}}{A}(t)$ and $\R{G}{A}(t)$ take the form of prescribed time-varying functions.

Given the aforementioned attitude parameterisation, the following quantities may now be related directly to the Euler angles $\theta$, $\psi$ and $\phi$.
\\ \\
{\bf Curvature:}
\begin{flalign}
  &\mb \kappa_x \\ \kappa_y \\ \kappa_z \me = \svec{R}{E}
  \begin{pmatrix}
    \psi^\prime\ct\sp+\theta^\prime\cp
    \\
    \phi^\prime-\psi^\prime\st\\
   -\psi^\prime\ct\cp+\theta^\prime\sp
  \end{pmatrix}\hs{20}
\end{flalign}
{\bf Spin:}
\begin{flalign}
  \svec{\delta\vartheta}{[A]} &= \svec{R}{E}\left(\delta\theta\mb\cs\\-\ss\\0\me
    + \delta\psi\mb0\\0\\-1\me+\delta\phi \: \svec{y}{E}\right)
  \label{eq:delta vartheta} \\
  \svec{\ddot\vartheta}{[A]}
    &= \svec{R}{E}\left(\ddot\theta\mb\cs\\-\ss\\0\me + \dot\theta\dot\psi\mb-\ss\\-\cs\\0\me +
                  \ddot\psi\mb0\\0\\-1\me+\ddot\phi \: \svec{y}{E}+\dot\phi \: \svec{\dot y}{E}\right)
  \label{eq:ddot vartheta}
\end{flalign}
\\
In summary, the variational terms of the problem have now effectively been cast in the form
\begin{flalign}
 \vec{\delta\Gamma}(\vec{\zeta},\vec{\zeta^\prime},\vec{\delta\zeta},\vec{\delta\zeta^\prime})\;,\hs{3}
 \vec{\delta\vartheta}(\vec{\zeta},\vec{\zeta^\prime},\vec{\delta\zeta},\vec{\delta\zeta^\prime})\;,\hs{3}
 \vec{\delta\xi}(\vec{\zeta},\vec{\zeta^\prime},\vec{\delta\zeta},\vec{\delta\zeta^\prime})\;,
\end{flalign}
parameterised by the components of the kinematic vector $\vec\zeta$ where
\begin{flalign}
    \vec{\zeta}(s,t)
        &= \mb \tau_x(s,t) \;,\; \varepsilon(s,t) \;,\; \tau_z(s,t) \;,\; \vec{p(s,t)^T} \me^T \notag\\
        &= \mb \tau_x(s,t) \;,\; \varepsilon(s,t) \;,\; \tau_z(s,t) \;,\; \theta(s,t) \;,\; \psi(s,t) \;,\; \phi(s,t) \me^T .
\end{flalign}

\subsection{Generalised Coordinates}\label{sec: Generalised Coordinates}
In line with d'Alembert's principle, one may state that the dynamic evolution of the system will proceed such that the virtual work contributions of equations (\ref{eq:delta W_K}), (\ref{eq:delta W_C}), (\ref{eq:delta W_F}), (\ref{eq:delta W_M}) and (\ref{eq:delta W_T}) will sum to zero for all admissible variations; thus
\begin{equation}
    \delta W_K + \delta W_C + \delta W_F + \delta W_M + \delta W_T = 0\;.
    \label{eq:dAlembert_variation}
\end{equation}
It has been shown that all of these virtual work contributions may be expressed in terms of the kinematic vector $\vec\zeta$. In order to construct the temporal ordinary differential equations capturing the system dynamics, the components of $\vec\zeta$ are written in the typical $s,t$ separable form as a summation of shape functions
\begin{flalign}
    \zeta_j(s,t) = \sum_{k}B_{jk}(s)\,q_{jk}(t)+\zeta_{j0}(s)\hs{1.5}:\hs{6}
    \mathrm{e.g.}\hs{4}\theta(s,t) = \sum_{k}B_{4k}(s)\,q_{4k}(t)+\theta_0(s)
    \label{eq:zeta_shape_function_summation}
\end{flalign}
where $\vec{B}$ is the set of shape functions approximating the deflected state of the flexible beam continuum and $\zeta_{j0}(s)$ denotes some optional reference distribution for each kinematic parameter from which the shape functions deform the system. $q_{jk}$ are the time dependent generalised coordinates that will form the state vector of this beam formulation. Note that the tensor notation is adopted throughout the remainder of this paper wherein one is to sum over all values of any index that appears {\bf only} in multiplicative pairs. Hence, equation (\ref{eq:zeta_shape_function_summation}) is written equivalently as
\begin{flalign}
    \zeta_j(s,t) = B_{jk}(s)\,q_{jk}(t)+\zeta_{j0}(s)\;.
    \label{eq:zeta_form}
\end{flalign}


\subsection{Equations of Motion}\label{sec: Equations of Motion}
The partial derivatives of (\ref{eq:dAlembert_variation}) are now taken with respect to each of the generalised coordinates $q_{jk}$
\begin{flalign}
  \pdiff{W_K}{q_{jk}} + \pdiff{W_C}{q_{jk}} + \pdiff{W_F}{q_{jk}} + \pdiff{W_M}{q_{jk}} + \pdiff{W_T}{q_{jk}} = 0\;.
  \label{eq:Lagrange system dynamic}
\end{flalign}
These terms are broken down into their constituent parts via chain differentiation in appendix section \nolink{\ref{Asec: appendix}}\ref{Asec: work derivatives}.

Writing the above equation (\ref{eq:Lagrange system dynamic}) as a matrix system yields
\begin{flalign}
\vec{\pdiff{W_K}{q}}+\vec{\pdiff{W_C}{q}}+\vec{\pdiff{W_F}{q}}
                           +\vec{\pdiff{W_M}{q}}+\vec{\pdiff{W_T}{q}} = \vec{0}
\end{flalign}
where the differential matrix operator
\begin{flalign}
  \vec{\pdiff{}{q}} =
  \left( \pdiff{}{q_{11}} ,\,...\:,
      \pdiff{}{q_{6k}} \right)^T .
\end{flalign}
To facilitate solution of this system, the acceleration terms in the above formulation are rearranged, factoring out any expressions that include the 2nd order state derivative $\vec{\ddot q}$. This rearrangement concerns the terms $\svec{\ddot\vartheta}{[G]}$ and $\svec{\ddot\Gamma}{m[G]}$ and is detailed in appendix section \nolink{\ref{Asec: appendix}}\ref{Asec: Re-arrangement of Kinetic Terms}. Consequently, the kinetic virtual work $\vec{\pdiff{W_T}{q}}$ may be split into the following parts (example matrix dimensions are also provided where $Q$ is equal to the number of states in $\vec{q}$).\\
\begin{flalign}
  \underset{Q\times 1}{\vec{\pdiff{W_T}{q}}} =
  &-\left[
    \intt{0}{L}m(s)\underset{Q\times 3}{\vec{\left(\pdiff{r_{m[G]}}{q}\right)^T}}
                   \underset{3\times Q}{\vec{\left(\pdiff{r_{m[G]}}{q}\right)}}
             + \underset{Q\times 3}{\vec{\left(\pdiff{\vartheta_{[G]}}{q}\right)^T}}
                   \underset{3\times 3}{\svec{\E}{[G]}\vphantom{\int}}
                   \underset{3\times 3}{\svec{I}{\vartheta [I]}\vphantom{\int}}
                   \underset{3\times 3}{\vec{\E_{[G]}^T}\vphantom{\int}}
               \underset{3\times Q}{\vec{\left(\pdiff{\vartheta_{[G]}}{q}\right)}}\d s
  \right]\underset{Q\times 1}{\vec{\ddot q}\vphantom{\int}} \notag\\
  &-\left[
    \intt{0}{L}m(s)\underset{Q\times 3}{\vec{\left(\pdiff{r_{m[G]}}{q}\right)^T}}
                       \underset{3\times 1}{\vec{\ddot r_{m[G]}^*}\vphantom{\int}}\right.\notag\\[-2mm]
                 &\quad+ \left.\underset{Q\times 3}{\vec{\left(\pdiff{\vartheta_{[G]}}{q}\right)^T}}
                       \underset{3\times 3}{\svec{\dot\E}{[G]}\vphantom{\int}}
                       \underset{3\times 3}{\svec{I}{\vartheta [I]}\vphantom{\int}}
                       \underset{3\times 3}{\vec{\E_{[G]}^T}\vphantom{\int}}
                       \underset{3\times 1}{\vec{\dot\vartheta_{[G]}}\vphantom{\int}}
                 + \underset{Q\times 3}{\vec{\left(\pdiff{\vartheta_{[G]}}{q}\right)^T}}
                       \underset{3\times 3}{\svec{\E}{[G]}\vphantom{\int}}
                       \underset{3\times 3}{\svec{I}{\vartheta [I]}\vphantom{\int}}
                       \underset{3\times 3}{\vec{\E_{[G]}^T}\vphantom{\int}}
                       \underset{3\times 1}{\vec{\ddot\vartheta_{[G]}^*}\vphantom{\int}}\d s
  \right] \notag\\
  = &-\vec{M}\vec{\ddot q} - \vec{w}
  \label{eq:W_T_rearrangement}
\end{flalign}
where
\begin{flalign}
  \vec{\ddot\vartheta_{[G]}^*} &= \svec{\dot{\bar\Omega}}{A[G]}
      + \svec{\bar\Omega}{A[G]}\times\svec{\dot\vartheta}{[G]} + \R{G}{A}\svec{R}{E}
        \left(\dot\theta\dot\psi\mb-\ss\\-\cs\\0\me + \dot\phi \: \svec{\dot y}{E}\right)\;,\\
  \vec{\ddot r_{m[G]}^*} \;&=\; \vec{{\ddot{\bar r}}_{A[G]}} + \vec{{\ddot\E}_{[G]}^*}\mb a_m \\ 0 \\ c_m \me
    + \intt{0}{s} \vec{{\ddot\E}_{[G]}^*} \mb\tau_x\\1+\varepsilon\\\tau_z\me + 2\svec{\dot\E}{[G]} \mb\dot\tau_x\\\dot\varepsilon\\\dot\tau_z\me \d \tilde s \;,\\[4mm]
  \vec{\ddot\E_{[G]}^*} &= \vec{\ddot\vartheta_{[G]}^*}\times\svec{\E}{[G]} + \svec{\dot\vartheta}{[G]}\times\svec{\dot\vartheta}{[G]}\times\svec{\E}{[G]}\;.
\end{flalign}
Here, $\svec{\bar\Omega}{A[G]}$ is the angular velocity of the coordinate system ${\bf A}$.

Given this rearrangement of the dynamic terms, the final equations of motion for the nonlinear beam take the form of the ODE system:
\begin{flalign}
  \boxed{
  \vec{M\ddot q} = -\vec{w}+\vec{\pdiff{W_K}{q}}+\vec{\pdiff{W_C}{q}}
                           +\vec{\pdiff{W_F}{q}}+\vec{\pdiff{W_M}{q}}}\label{eq:EoM}
\end{flalign}
$\vec{M}$ and $\vec{w}$ are given by equation (\ref{eq:W_T_rearrangement}); the remaining terms $\vec{\delta{W_K}/\delta{q}}$, $\vec{\delta{W_C}/\delta{q}}$, $\vec{\delta{W_F}/\delta{q}}$ and $\vec{\delta{W_M}/\delta{q}}$ are as defined by (\ref{eq:pdiff{W_K}{q}}), (\ref{eq:pdiff{W_C}{q}}), (\ref{eq:pdiff{W_F}{q}}) and (\ref{eq:pdiff{W_M}{q}}), respectively, with constituent relations feeding into these terms as detailed in sections \nolink{\ref{Asec: appendix}}\ref{Asec: zeta derivatives}--\nolink{\ref{Asec: appendix}}\ref{Asec: Gamma derivatives}.


\section{Shape Function Basis}\label{sec: Shape Function Basis}
In order to implement the formulation detailed in the preceding section one must choose a suitable set (or sets) of shape functions from which approximations of the components of $\vec{\zeta}$ can be constructed. The choice of such a set is quite general, requiring only satisfaction of the essential kinematic boundary conditions of the problem. In the context of this formulation, these take the form of prescribed boundary values applied to a subset of $\vec\zeta$ components. Considering equation (\ref{eq:zeta_form}) one may write this condition
\begin{flalign}
  \zeta_j(\bar{S},t) &= B_{jk}(\bar{S})\,q_{jk}(t)+\zeta_{j0}(\bar{S}) = \bar{\zeta_j} \hs{20} \bar{S}\in\{0,L\} \notag\\
  &\Rightarrow\hs{2} B_{jk}(\bar{S}) = 0\hs{2},\hs{2}\zeta_{j0}(\bar{S}) = \bar{\zeta_j}
  \label{eq:kinematicBC}
\end{flalign}
where the value $\bar{\zeta_j}$ is constant. This condition holds for any constrained component of $\vec\zeta$. It is noted here that the reference system ${\bf A}$, in which this condition is cast, is a non-inertial system and thus (\ref{eq:kinematicBC}) still admits arbitrary translation and rotation of the beam in the global coordinate system G (see for example the spinning beam test case of section \nolink{\ref{sec: Numerical Test Cases}}\ref{sec: Rotating Pre-Curved Beam} ).

\begin{figure}[t]
\centering
\includegraphics[width=\linewidth,clip=true,trim=20mm 113mm 20mm 113mm]{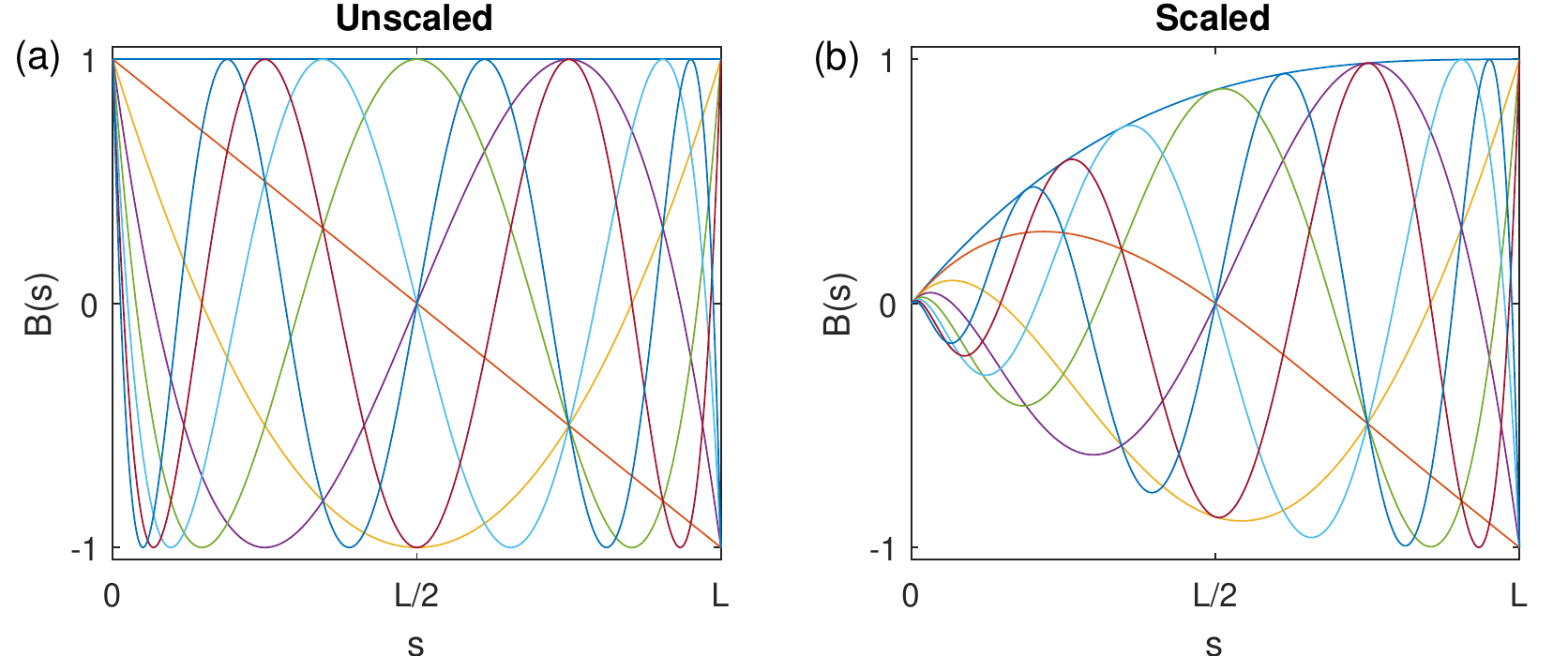}
\caption{The first eight functions of the candidate shape set based on Chebyshev polynomials of the first kind. Panels (a) depicts the pre-multiplied set; panel (b) depict the scaled set following application of the envelope function $E(s)$.}
\label{fig:chebyshev_set}
\end{figure}

A suitable set of shape functions satisfying (\ref{eq:kinematicBC}) is depicted in figure \ref{fig:chebyshev_set}(a); these functions are based on shifted and reversed Chebyshev polynomials of the 1st kind. In panel (b), the same set is depicted following multiplication by a scaling function $E(s)$ to bring the left side to zero (similar such scaling may be applied to the right hand side for tip boundary conditions). This scaling procedure is applied throughout the examples of this study to constrain the applicable components of $\vec{\zeta}$. A recursive generating relation for this set of functions $y_n(x)$ is given by
\begin{flalign}
    &B_n(x) = E(s)C_n(x(s)) & \label{eq:chebyshev_set} \\
    &x(s) = -2(s/L)+1 & s \in [0,L] \notag\\
    &E(s) = (s/L)^3-3(s/L)^2+3(s/L) & \notag\\
    &C_1(x) = 1 & \notag\\
    &C_2(x) = x & \notag\\
    &C_n(x) = 2xC_{n-1}(x)-C_{n-2}(x) & n \in \{1,2,\dots,N\} \notag
\end{flalign}
Note that the use of the scaling function $E(s)$ retains orthogonality of the underlying Chebyshev set, however this is not a required property; nor must each component of $\vec\zeta$ draw from the same shape function basis. Furthermore because of the use of this scaling function, it becomes trivial to satisfy the boundary condition (\ref{eq:kinematicBC}) for other arbitrary shape sets. However, the relative merits of different candidate shape bases are not pursued further in this study; rather the intention is to show that for the general set (\ref{eq:chebyshev_set}) satisfying the minimal essential boundary condition of the problem (\ref{eq:kinematicBC}), the various test cases detailed in section \ref{sec: Numerical Test Cases} can be efficiently treated (note also that throughout the cases of this paper, $\zeta_{j0}(s) = 0$). For similar reasons the properties of orthogonality and natural load dependent constraints are not discussed further here.

\begin{figure}[t]
\centering
\includegraphics[width=\linewidth,clip=true,trim=0mm 0mm 0mm 0mm]{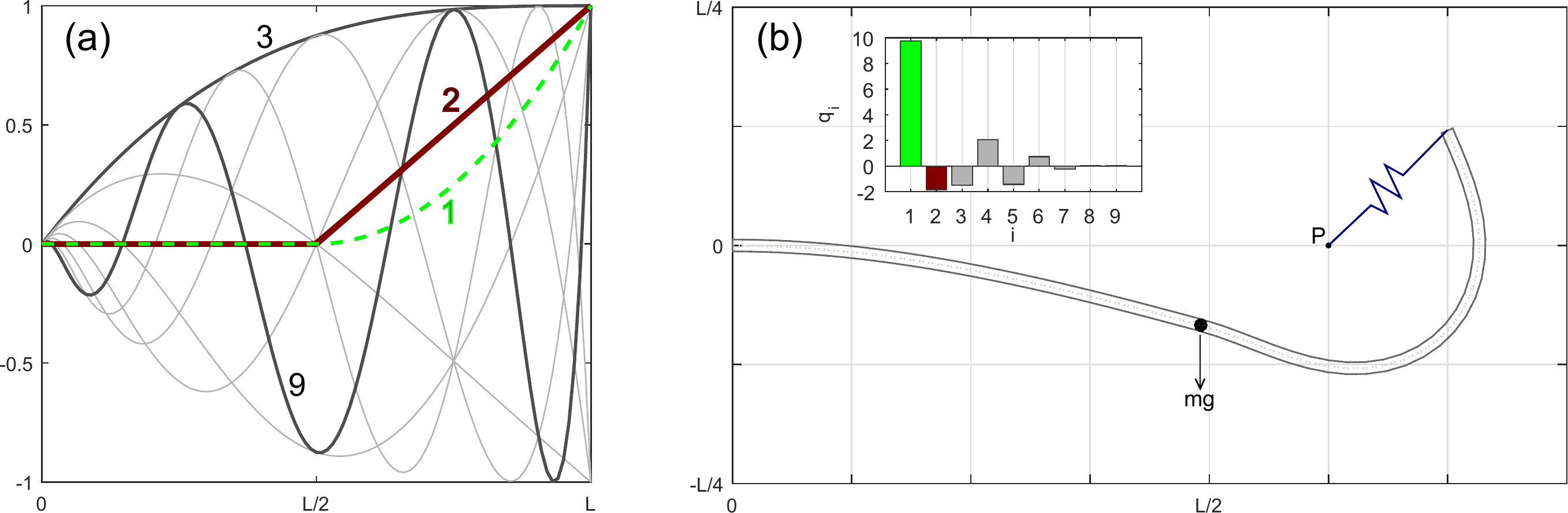}
\caption{An example non-smooth beam problem. Panel (a) gives the modified shape function set used to embed the stiffness and mass discontinuities in the numerical problem and panel (b) depicts the predicted static deflection obtained as a consequence of the mass distribution and tip spring.}
\label{fig:discontinuous_example}
\end{figure}
Before proceeding, it is noted that no condition on the smoothness or continuity of these shape functions has thus far been imposed; indeed any discontinuities of these sets may be treated via appropriate placement of the integration bounds of the weak formulation (\ref{eq:EoM}). Thus, discontinuous features of the physical problem may be treated by embedding the same discontinuities in the approximating shape basis. To illustrate this, consider the modelling of a straight horizontal cantilever beam with the following characteristics: Firstly, the inner half running between the root and mid point is assigned a much greater stiffness than the outer half from midpoint to tip; secondly, a point mass is fixed to the midpoint of the beam at the boundary of these two sections. One (basic) way of embedding these discontinuities into the underlying shape function set is to furnish the existing smooth functions with two additional non-smooth functions with appropriate mid-span $C^1$ and $C^2$ discontinuities; an example Chebyshev derived set, extended in this fashion, is depicted in figure \ref{fig:discontinuous_example}(a) with the additional $C^1$ and $C^2$ discontinuous functions given by the dashed line and thicker line labelled 1 and 2 respectively (the remaining Chebyshev functions are indicated in increasing order by the grey curves labelled 3--9). In panel (b) the treatment of an example static test is shown; here a spring is connected between the beam tip and point `P'. The combined effects of self weight and tip force induced by the spring cause the beam to assume the indicated shape with a clearly visible larger deformation of the more flexible outer section. The contributions of each of the shape functions to this static solution are given by the inset bar diagram in panel (b); the index of each bar refers to the corresponding shape function labelled in panel (a). Note that a more sophisticated tailoring of the underlying shape set may be employed in the efficient treatment these or other such localised characteristics, however, such a discussion is not pursued further in this study.

It is pointed out here that the treatment of such discrete features also extends to geometric discontinuities in the structure such as for a kinked beam or framed structure (see examples in \cite{Santos2011_etal,Crisfield1990,Simo_Pt2,Pai2007,Ibrahimbegovic2000}).
In such cases, the discontinuity must also be represented in the kinematic description of the system. There are multiple entry points in the above formulation where this discontinuous information may be embedded. For example, one possibility is to define a generic mapping $\svec{T}{D}$ from the underlying Euler-referenced triad system  $\svec{\E}{[E]}$ to the structure-aligned intrinsic triad system $\svec{\E}{[A]}$ used in calculating the global reference line $\svec{\Gamma}{[A]}$, i.e.
\begin{flalign*}
\svec{\E}{[A]} = \svec{R}{E}\mb\svec{x}{E} \mid \svec{y}{E} \mid \svec{z}{E}\me\vec{R_E^T}\svec{T}{D}\;.
\end{flalign*}
Another possibility is to directly incorporate appropriate Dirac-delta functions into the initial curvature distribution $\svec{\xi}{0}$ to built up the discontinuous global geometry. These potential extensions of the formulation and their relative merits to the treatment of frame structures and other discontinuous geometries lies outside of the scope of this study.

Finally, it is worth mentioning that, alternatively, for the above examples one may of course treat smooth sub-domains of the problem breaking the discontinuities across designated element boundaries. Combined with a suitable choice of finite-element type basis set one approaches the more typical treatment of other element-based geometrically exact methods. However, in doing so one invariably moves away from the minimal state representation sought in this study; furthermore one notes that although smooth problems are exclusively treated in the examples of this study, the arbitrary smooth variation of physical parameters as displayed in the example of section \nolink{\ref{sec: Numerical Test Cases}}\ref{sec: Wind Turbine Blade} is successfully treated in a geometrically exact manner without subdivision of the problem; this feature is novel to this \methodName formulation.

\subsubsection*{A Note on Numerical Integration}

Consider the numerical integration of the \methodName formulation. Recall the previously derived equation of motion
\begin{flalign}
  &\vec{M\ddot q} = -\vec{w}+\vec{\pdiff{W_K}{q}}+\vec{\pdiff{W_C}{q}}
                           +\vec{\pdiff{W_F}{q}}+\vec{\pdiff{W_M}{q}} \notag \\
  \mathrm{or}\hs{15}&\vec{M(q,B(s))\;\ddot q} = \vec{f(q,\dot q,B(s))}\;.
  \label{eq:EoM_form}
\end{flalign}
Since the choice of shape functions $\vec{B(s)}$ is arbitrary, the constituent integral terms of $\vec M$ and $\vec f$ cannot typically be treated analytically. Specifically, the terms $\vec\Gamma$, $\vec{\dot\Gamma}$, $\vec{\pdiff{\Gamma}{q}}$, $\vec M$, $\vec w$, $\vec{\pdiff{W_K}{q}}$, $\vec{\pdiff{W_C}{q}}$, 
$\vec{\pdiff{W_F}{q}}$ and $\vec{\pdiff{W_M}{q}}$ are integrated numerically over $s\in[O,L]$ when evaluating the equation of motion at a particular time step. Here, numerical evaluation of these terms is accomplished using quadratic interpolation. Noting that the full system is stiff, one may compute a dynamic trajectory in time using any stiff 1st or 2nd order solver capable of treating systems of the form (\ref{eq:EoM_form}).
Since this method is implemented in Matlab, the in-built variable-step, variable-order stiff ODE solver `ODE15s' (see \cite{Shampine1997}) is applied to the 1st order form of (\ref{eq:EoM_form}).

\section{Numerical Test Cases}\label{sec: Numerical Test Cases}

A variety of test cases, comprising benchmark tests from the literature as well as some further examples, are now presented. Each case targets particular characteristics of the geometrically nonlinear flexible beam problem. In addition to published results, the responses generated by MSC Nastran and the intrinsic beam formulation of Hodges \cite{Hodges1990,Hodges1996} are in places used to validate the \methodName approach proposed in this paper. For all the cases the Chebyshev polynomial shape set detailed in section \ref{sec: Shape Function Basis} are used in approximating $\tau_x(s)$, $\varepsilon(s)$, $\tau_z(s)$ (figure \ref{fig:chebyshev_set}(a)) and $\theta(s)$, $\psi(s)$, $\phi(s)$ (figure \ref{fig:chebyshev_set}(b)).

\subsection{Large static deformation}
\begin{figure}[t]
\centering
\begin{minipage}{.49\textwidth}
\raisebox{4cm}{(a)}
\includegraphics[width=.95\textwidth,clip=true,trim=16mm 0mm 8mm 0mm]{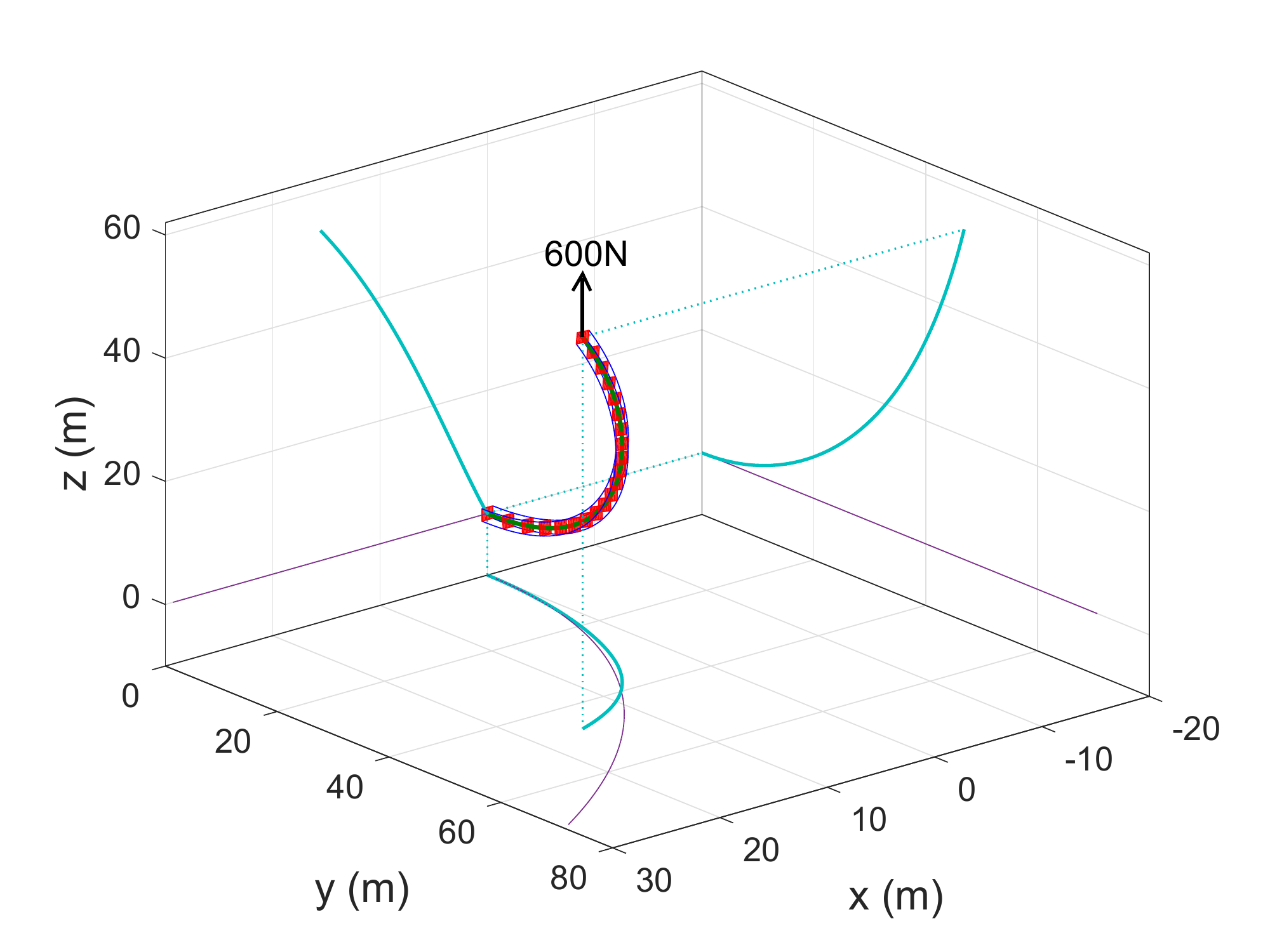}
\end{minipage}
\begin{minipage}{.49\textwidth}
\raisebox{4cm}{(b)}
\includegraphics[width=.95\textwidth,clip=true,trim=16mm 0mm 8mm 0mm]{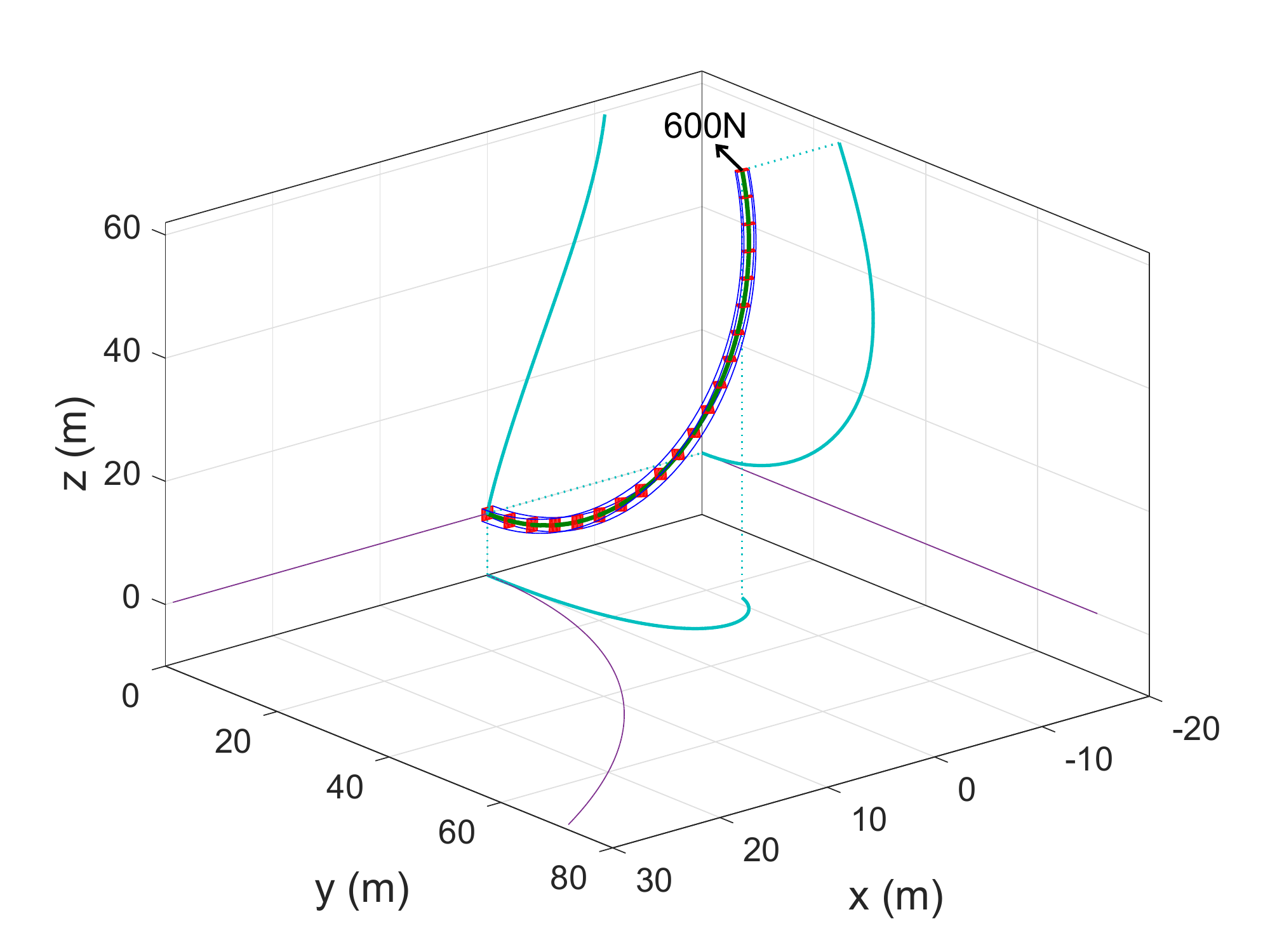}
\end{minipage}
\caption{Beam deformation given (a) a 600N vertical tip load (b) a 600N follower tip load}
\label{fig: 45deg_pre_curve}
\end{figure}
The first test case considered is well documented in the literature and treats the large deformation of a pre-curved beam. The initial shape of the beam forms 1/8th of a circle of radius 100m in the horizontal (x,y) plane (see the thin curves in figure \ref{fig: 45deg_pre_curve}). The beam has a 1m square cross section, a Young's modulus of $E = 10^7 \mathrm{N/m^2}$ and a Poisson's ratio of zero (thus $EI_{xx} = EI_{zz} = 25/3\;\mathrm{Nm^2}$, $GJ = 7.02885\e5\;\mathrm{Nm/rad}$); gravitational acceleration is not considered. The system is solved statically using the Euler mapping
\begin{equation}
  \svec{R}{E} = \mb0&0&-1\\0&1&0\\1&0&0\me
\end{equation}
moving the singular points away from vertical (all other test cases in this study are treated with $\svec{R}{E}$ equal to the 3$\times$3 identity matrix).
Two load cases are shown in figure \ref{fig: 45deg_pre_curve} each involving the application of a 600N tip load. In panel (a) the tip load is applied in the vertical z direction; in panel (b) the same load is applied as a follower force (i.e. remaining parallel to $\svec{e}{z}(L)$). In table \ref{table: 45deg_pre_curve_tip_deflections} the static tip deflection in the global coordinate system is stated as predicted by equivalent tests from a number of published sources. Supplementary results using Nastran and the intrinsic beam formulation are also displayed and a good agreement is observed.
\begin{table}[h]
\vspace{-0.3cm}
\centering
\begin{tabular}{ |l|c|c| }
  \hline
  Source & Tip Position (m) $\mathrm{F_{vertical}}$ & Tip Position (m) $\mathrm{F_{follower}}$ \\
  \hline
    Bathe and Bolourchi \cite{Bathe1979} & (15.9$\msz{}{0}$, 47.2$\msz{}{0}$, 53.4$\msz{}{0}$ & -- \\
    Simo and Vu-Quoc \cite{Simo_Pt2} & (15.79, 47.23, 53.37) & -- \\
    G\'{e}radin and Cardona \cite{Geradin2001} & (15.55, 47.04, 53.50) & -- \\
    Li and Vu-Quoc \cite{Li2010} & (15.54, 46.85, 53.64) & -- \\
    Crisfield \cite{Crisfield1990} & (15.61, 46.84, 53.71) & -- \\
    MSC Nastran & (15.56, 46.89, 53.60) & (-10.92, 24.53, 59.42) \\
    Intrinsic Beam & (15.56, 46.90, 53.60) & (-10.93, 24.55, 59.41) \\
    \MethodName \raisebox{3.4mm}{} & (15.55, 46.90, 53.60) & (-10.95, 24.54, 59.41) \\
 \hline
\end{tabular}
\caption{Comparison of predicted tip positions for the curved beam static test case.}
\vspace{-0.5cm}
\label{table: 45deg_pre_curve_tip_deflections}
\end{table}
\subsubsection{Problem Size}

To assess the number of states required for the various methods, consider the following convergence criteria on the required number of states $n_q^*$
\begin{equation}
n_q^*\leq \forall n_q : \{n_q\in\mathbb{N}\hs{2}|\hs{2}|\svec{\Gamma}{tip}(n_q) - \svec{\Gamma}{tip}(2n_q)|<0.001\}
\label{eq:Convergence criterion}
\end{equation}
i.e. doubling the number of states alters the predicted tip deflection by less than 1mm.
For the \methodName method, this doubling is applied to the discretisation for each $\vec\zeta$ component; for the special case of zero component shapes, one shape is added rather than doubling the set size.

Using (\ref{eq:Convergence criterion}), the total number of states required by the \methodName method, the intrinsic beam formulation and MSC Nastran is shown in table \ref{table: 45deg_pre_curve_required_DoFs}; results are given for both the tip vertical (figure \ref{fig: 45deg_pre_curve}(a)) and tip follower (figure \ref{fig: 45deg_pre_curve}(b)) load cases. For the current formulation the number of shape functions assigned to $\tau_x(s)$, $\epsilon(s)$, $\tau_z(s)$, $\theta(s)$, $\psi(s)$ and $\phi(s)$ is indicated in brackets respectively. In addition, table \ref{table: 45deg_pre_curve_required_DoFs} also provides a rough comparison of the required computational effort in treating these systems. To achieve this the follower and vertical load cases are run as dynamic simulations over a period of 60 seconds. A large stiffness proportional damping is used ($\svec{F}{K} = -\vec{\tilde K}\vec{\Delta\xi}$, $\svec{F}{C} = -\vec{\tilde K}\vec{\Delta\dot\xi}$) such that the responses are settled to their respective equilibria by the end of the 60 second simulation. The \methodName and intrinsic beam methods are run in MATLAB using the stiff ODE solver `ODE15s' \cite{Shampine1997}; no Jacobian information is fed to the solver for the former method.
Nastran is run using its in-built SOL400 solution routine with adaptive time stepping.
Due to the large deformations involved in this test case it proved difficult to obtain an accurate Nastran transient response for the follower 600N tip load. Hence only the Nastran benchmark time based upon the vertical load case is provided in table \ref{table: 45deg_pre_curve_required_DoFs}.
Because of the differences in the numerical integration schemes only a crude comparison of computational speed can be inferred here.
Looking to the results of table \ref{table: 45deg_pre_curve_required_DoFs} the \methodName formulation required an order of magnitude fewer states, accompanied by a significantly reduced simulation time when compared to the other methods.
Of particular note is that no shear or extensional states were required by the \methodName formulation in meeting the convergence criterion in this example. Indeed, this is the case for a majority of the example problems of this study wherein capturing large nonlinear deformations results in the treatment of flexible and slender structures where shear effects are not significant. One advantage of this formulation is that the shear and extensional compliances can easily be removed by simply not assigning shapes to $\tau_x(s)$, $\epsilon(s)$ and $\tau_z(s)$. Note however that the shallow arch example of the following section introduces a test case for which these additional shear/extensional flexibilities do prove influential on the resulting structural response.
\begin{table}[h]
\vspace{-0.3cm}
\centering
\begin{tabular}{ |l|c|c|c| }
  \hline
  Method and Load Case & Problem Discretisation & Static DoFs & Dynamic Solution \\
  & & & Time (seconds) \\
  \hline
  Nastran (vertical) & 22 elements & 132 & 40.1 \\
  Nastran (follower) & 60 elements & 360 & --- \\
  Intrinsic Beam (vertical) & 31 elements & 186 & 132 \\
  Intrinsic Beam (follower) & 69 elements & 414 & 760 \\
  \MethodName (vertical) \raisebox{3.4mm}{} & 11 shapes $(0,0,0,4,4,3)$ & 11 & 3.26 \\ \rowcolor{cLGy}
  \MethodName (follower)  \raisebox{3.4mm}{} & 15 shapes $(0,0,0,5,5,5)$ & 15 & 4.25 \\
  \hline
\end{tabular}
\caption{Static degrees of freedom required in satisfying the convergence criterion (\ref{eq:Convergence criterion}).}
\vspace{-0.5cm}
\label{table: 45deg_pre_curve_required_DoFs}
\end{table}

\subsection{Shallow Arch Example}\label{sec: Shallow Arch Example}

This test case focuses on the response of a pin-jointed shallow arch subject to a distributed radial load. The motivating features of this test case are two-fold. Firstly, the treatment of non-cantilever boundary conditions provides a further example of the construction of equivalent kinematic boundary conditions in the context of this formulation. Secondly, the loaded arch provides a system for which the shear/extension states prove influential on the observed response.

The shallow arch is based upon the pre-curved beam of the previous example. The cross sectional properties remain the same, however the radius of curvature is now reduced to 20m --- consequently reducing the beam length over the 45 degree arc. This curved beam is supported on two pin joints to form an un-stressed arch to which a radial load is applied. This problem is planar and oriented in the (y,z)--plane; thus $\psi(s) = \phi(s) = \tau_x(s) = 0$, a condition achieved simply by assigning no shape functions to the 1st, 5th and 6th components of $\vec{\zeta}$. The remaining $\theta(s)$, $\varepsilon(s)$ and $\tau_z(s)$ kinematic parameters are unconstrained at both boundaries of the arch, thus the unscaled shape set of figure \ref{fig:chebyshev_set}(a) is selected for these components. $\svec{\Gamma}{[G]}(0) = \svec{\Gamma}{0}$ and $\svec{\Gamma}{[G]}(L) = \svec{\Gamma}{L}$ correspond to the displacement constraints at each end of the arch. $\svec{\Gamma}{[G]}(0) = \svec{\Gamma}{0}$ is implicitly satisfied by equation (\ref{eq:gamma}) where $ \svec{\bar r}{A[G]} = \svec{\Gamma}{0}$; to constrain the right-hand boundary equation (\ref{eq:Lagrange system dynamic}) is furnished with the additional energy term
\begin{flalign}
  \delta W_\Upsilon = \Upsilon\,(\svec{\Gamma}{[G]}(L)- \svec{\Gamma}{L})\,\delta\svec{\Gamma}{[G]}(L)
\end{flalign}
where the multiplier $\Upsilon$ is given a sufficiently large value to constrain $\svec{\Gamma}{[G]}(L)$.

Application of a distributed radial load will cause this arch to buckle. Despite the simplicity of the configuration there are a number of distinct buckling behaviours that may be observed for this system. In \cite{Pi2007}, Pi \etal classify these as snap-through and bifurcation buckling modes, the latter of which is typically asymmetric in nature. The specific buckling characteristic observed is related to the non-dimensional `shallowness' parameter $\lambda$. For a rigid-pinned circular arch such as that treated in this example this parameter takes the form
\begin{flalign}
  \lambda = \frac{L^2\sqrt{A}} {4 R \sqrt{I_1}}
\end{flalign}
where $A$ is the cross section area of the beam, $I_1$ is the second moment of area and $R$ is the radius of curvature. This yields a lambda value of 10.68 for this test case.
\\
\begin{figure}[t]
\centering
\begin{minipage}{0.9\textwidth}
\includegraphics[width=.95\textwidth,clip=true,trim=18mm 105mm 18mm 108mm]{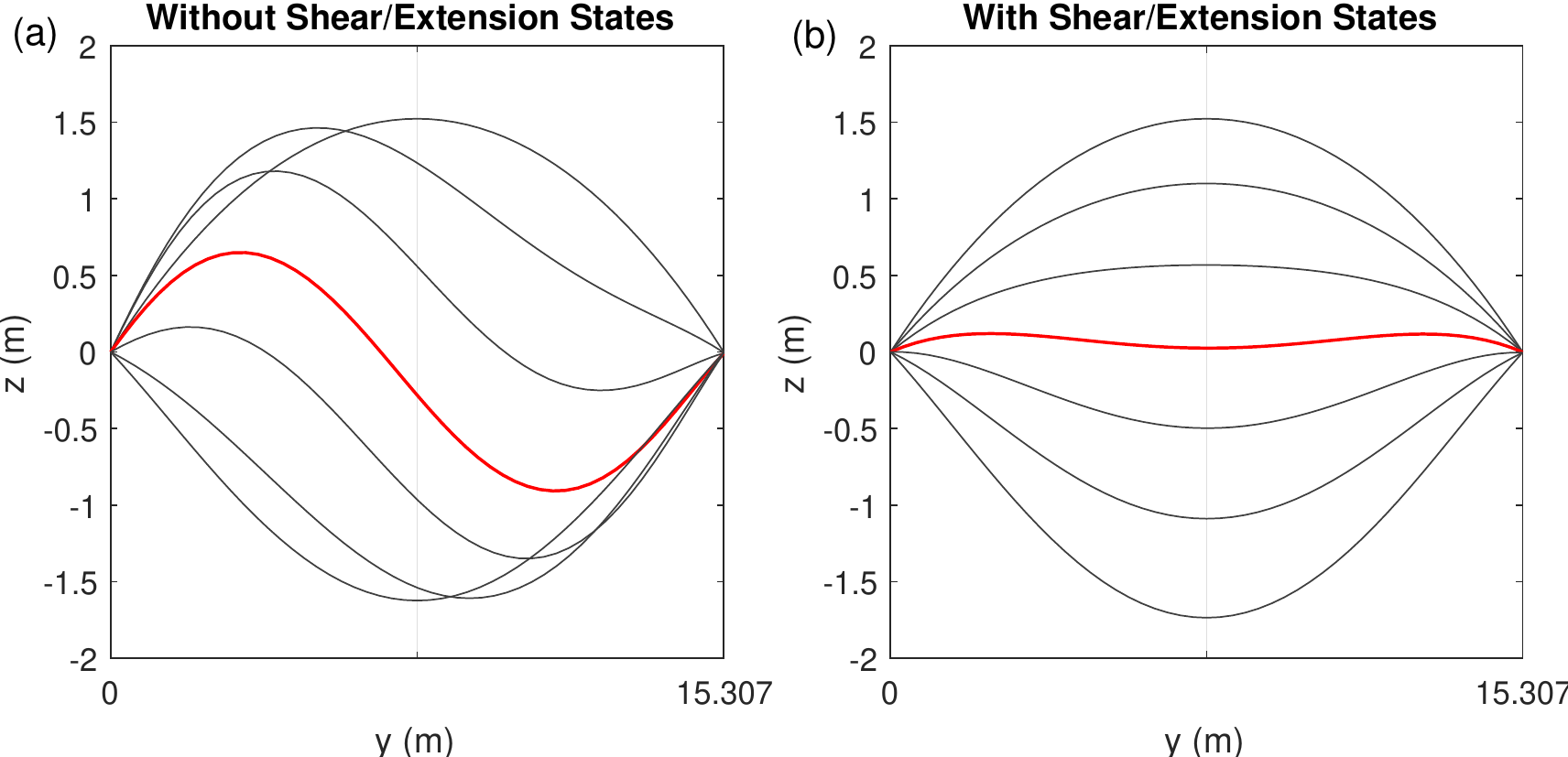}
\end{minipage}
\caption{Arch buckling following application of a distributed load. Panel (a) excludes shear and extensional states; panel (b) includes shear/extension. In both cases five shapes are used for each active component of $\vec\zeta(s)$.}
\label{fig: shallow_arch}
\end{figure}
\\
Looking now to figure \ref{fig: shallow_arch}, the shape of the shallow arch is shown following application of a radial load and snapshots of the deformation of the arch as it buckles are depicted. In panel (a) a distributed load of 7500 N/m is applied; no shear or extensional compliance is permitted and the arch is observed to buckle under an asymmetric bifurcation. This agrees with analytic results for a rigidly supported arch with $\lambda$ above the critical value of $\sim 9.2$. The effect of elastically compliant supports are also considered in \cite{Pi2007}. It is shown that as the flexibility of these supports is increased, the effective $\lambda$ is reduced, allowing for symmetric snap-through behaviour. Rather than directly modifying the support flexibility in this example, the addition of shear and extensional states are utilised to produce an alternative source of radial and axial compliance to the system. In figure \ref{fig: shallow_arch}(b), five shapes are assigned to each of the extensional $\varepsilon(s)$ and shear $\tau_z(s)$ parameters; the applied load is reduced to 5500 N/m in light of the increased flexibility of the arch. Looking at the snapshots of the arch deformation one observes that the addition of shear and extensional states is indeed sufficient to capture the existence of a symmetric snap-though mode.

\subsection{Rotating Pre-Curved Beam}\label{sec: Rotating Pre-Curved Beam}

The following two examples are based upon the test beam treated in Pai \cite{Pai2007}. The parameters are summarised in table \ref{table: Pai_beam_parameters}. For the first test case the beam is oriented in the global y direction ($\svec{e}{y[A]}(0) = (0,1,0)^T$) and assigned a constant pre curvature $\kappa_x = -(3\pi)/(18L)\;\;\mathrm{m^{-1}}$.

\begin{table}[h]
\vspace{-0.3cm}
\centering
\begin{tabular}{ |l|c|c| }
  \hline
  Parameter & Value & Units \\
  \hline
  Length & $0.479$ & m \\
  Width & $5.08\e{-2}$ & m \\
  Height & $4.5\e{-4}$ & m \\
  $EI_{xx}$ & $4.899184\e{-2}$ & $\mathrm{Nm^2}$ \\
  $EI_{zz}$ & 6.243471 & $\mathrm{Nm^2}$ \\
  $GJ$ & $7.164459\e{-2}$ & Nm/rad \\
  mass per length & $1.012698\e{-1}$ & kg/m \\
  torsional inertia per length & $2.178012\e{-1}$ & kg m \\
  \hline
\end{tabular}
\caption{Pai beam parameters from \cite{Pai2007}.}
\vspace{-0.5cm}
\label{table: Pai_beam_parameters}
\end{table}

\begin{figure}[t]
\centering
\includegraphics[width=1\textwidth,clip=true,trim= 66mm 110mm 225mm 17mm]{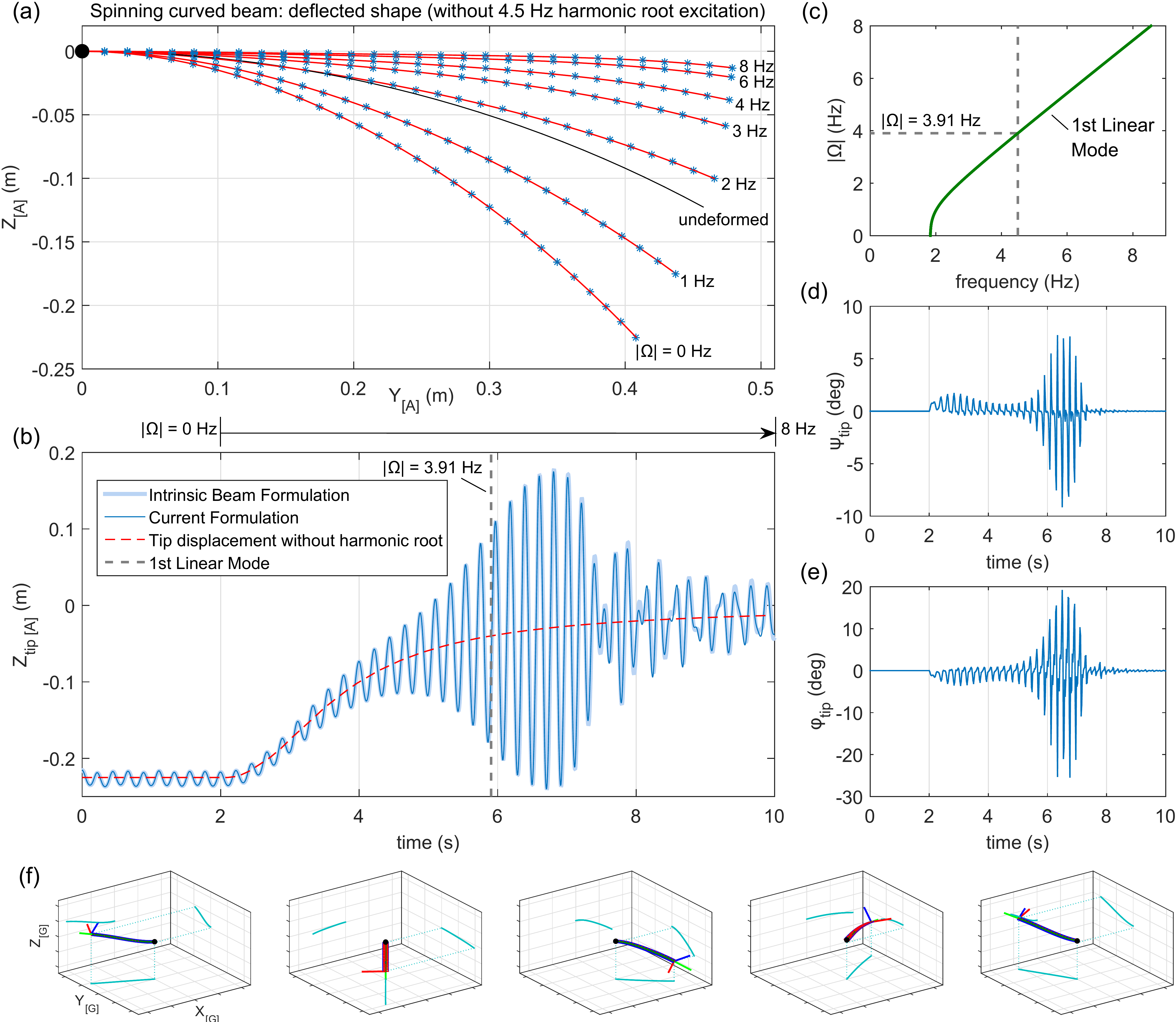}
\caption{Predicted responses for the pre-curved spinning beam test case. }
\label{fig: spinning_root_pre_curved}
\end{figure}

This pre-curved geometry is shown in panel (a) of figure \ref{fig: spinning_root_pre_curved} by the curve labelled `undeformed'. The additional results are obtained by first allowing the beam to deflect under its own weight (0~Hz curve). Then the root is spun in the global z direction at a number of angular velocities. The settled deformation as predicted by the shape function formulation is depicted by the labelled red lines for rotation speeds of 1, 2, 3, 4, 6 and 8 Hz. Corresponding data points from \cite{Pai2007} are overlaid onto these curves illustrating a very close agreement.
For the second phase of this test the root of the beam is fed a 4.5Hz harmonic excitation in the vertical direction of amplitude $\pm$2cm in addition to the applied angular rotation.
To help with visualisation the smaller plots at the bottom of figure \ref{fig: spinning_root_pre_curved} (panel (f)) illustrate the oscillating and spinning motion of this system in the global coordinate system ($X_{[G]},Y_{[G]},Z_{[G]}$).
A stiffness proportional, viscous damping law is assumed at 1\% of the structural stiffness; i.e. $\svec{F}{C}(s,\vec{\Delta\xi},\vec{\dot\xi}) = -0.01\vec{\tilde K(s)}\vec{\dot\xi(s)}$ in equation (\ref{eq:delta W_C}). The blue solid line in panel (b) shows the tip response for 10 seconds of this test case. At first the rotational velocity of the system is zero and thus the initial beam response is due purely to the root vertical oscillation. The settled response to this excitation is shown for the first 2 seconds of the time series. Then for $2\leq t\leq10$ the rotational velocity $|\svec{\Omega}{[G]}|$ is ramped linearly from 0 to 8 Hz. Over this time the beam tip continues to oscillate and remains centred about the dashed red line which corresponds to the static solution without the root vertical oscillation (compare with panel (a)). At the onset of the rotational sweep ($t=2$) one also notes both an out-of-plane and torsional component of the beam response (depicted in panels (d) and (e) by the non-zero $\psi(L)$ and $\phi(L)$ components of the response). This illustrates the expected gyroscopic effects resulting from large local rotational motion of the structure in the rotating beam reference frame.

Looking at these responses one notes that the amplitude of oscillation grows as the root angular velocity is increased reaching a maximum before dropping off again. Physically one is observing here the increasing centrifugal dynamic component having a stiffening effect on the flexible system and consequently increasing the frequencies at which the flexible modes reside (see for example the modelling studies of \cite{Berzeri2002,Fung1999,Yang2003} for treatments of the centrifugal stiffening effect in the context of simple rotating systems). This stiffening is illustrated in panel (c) where the relationship between the root angular velocity and the first modal linear frequency is depicted. For a rotational velocity of 3.91Hz --- occurring at $t=5.91$s in the simulated test case and indicated by the vertical line in panel (b) --- this modal frequency is equal to 4.5Hz and therefore equal to the frequency of root excitation of the system. Given a small amplitude excitation and sufficiently slow sweep the maximum response amplitude would be observed close to this angular velocity. However, for the level of harmonic excitation applied in this example the amplitude of the system response is sufficient to push into the nonlinear regime of the system dynamics and consequently the resonant frequency at the observed amplitudes will not coincide with this linear frequency; coupled with the relatively quick traversal through this resonant region one observes a delayed maximal response at a larger rotational frequency. To verify this case the same test was performed using the intrinsic beam model; the resulting tip deflection is also depicted in the background of panel (b) by the thicker background line. Here one sees that the obtained envelope of oscillation is almost identical between the two methods and that the time series match well, deviating only very slightly towards the end of the simulation.

\subsection{Harmonically Excited Vertical Cantilever}

\begin{figure}[t]
\centering
\includegraphics[width=1\textwidth,clip=true,trim=37mm 10mm 54mm 5mm]{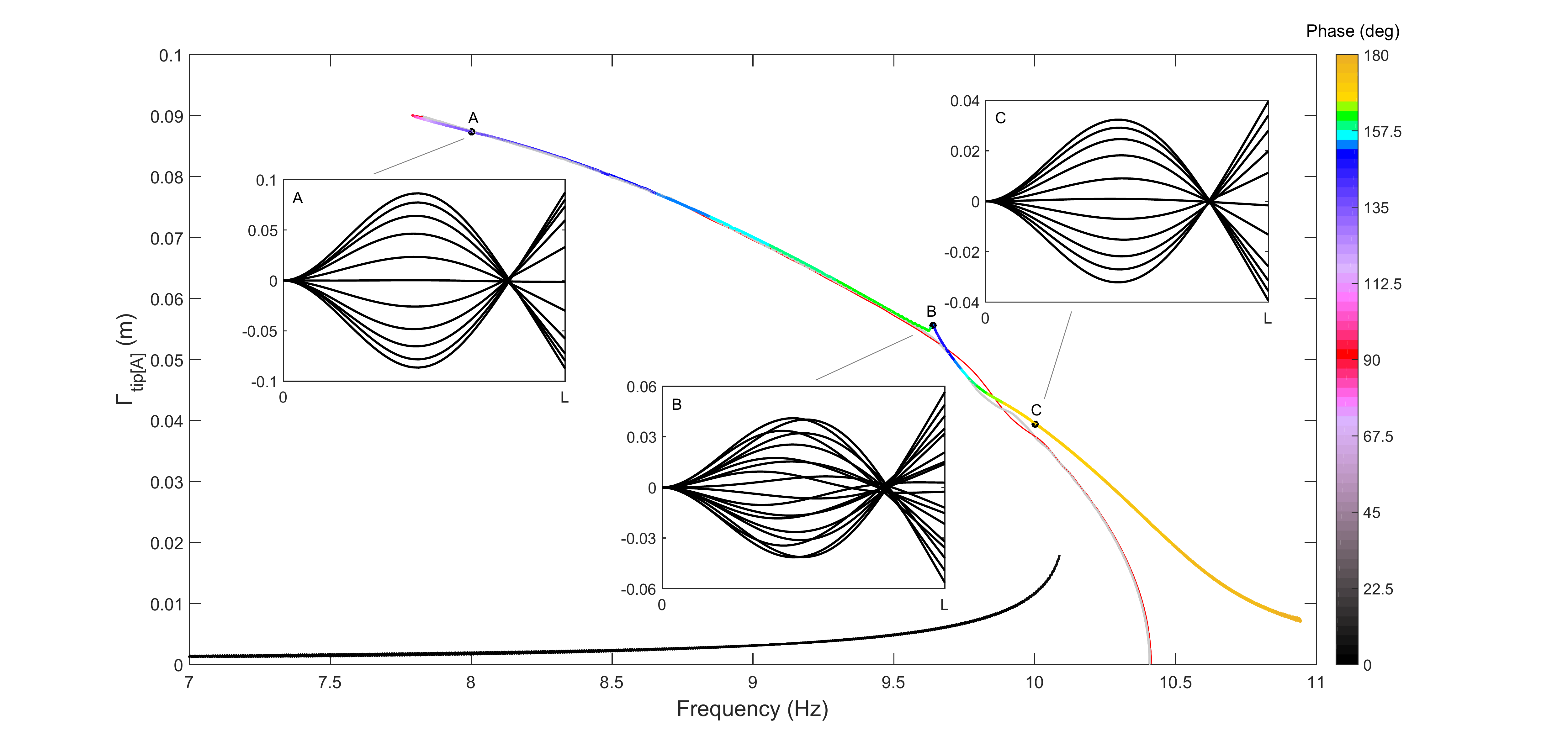}
\caption{Example frequency response diagram approximated via numerical simulation of the system for a 0.005\% stiffness proportional damping and $\pm$1mm base excitation. Estimates of the underlying nonlinear normal mode, as predicted by the current formulation and intrinsic beam model, are indicated by the thin red and thin grey curves, respectively. Snapshots of the beam oscillations are provided at points A (8Hz), B (9.64Hz) and C (10Hz) along the upper periodic branch.}
\label{fig: backbone}
\end{figure}

This next test case also concerns the beam detailed in the previous example. A harmonic root excitation is again fed to this system, this time in isolation. The beam configuration matches the corresponding experimental/numerical study conducted in \cite{Pai2007}. Specifically, the beam is oriented pointing vertically upwards and subject to its own weight. The transverse harmonic oscillation applied to the beam root acts in the $\pm\svec{e}{z}(0)$ direction and has an amplitude of $\pm$1mm. Structural damping is again stiffness proportional and set to a very light value of 0.005\% of the beam stiffness. Two slow frequency sweeps of the root harmonic excitation are performed over the range 7--11 Hz in both the forward and reverse directions. The amplitude of tip oscillation relative to the base position, as this excitation frequency is varied, is plotted by the thicker curves in figure \ref{fig: backbone}. These curves form approximations of the upper and lower stable periodic branches of the frequency response function for this level of root excitation and damping (the curves are clipped off at points where a loss of stability is observed). The colour at any location along these curves corresponds to the observed phase shift between the tip response and base excitation as indicated by the bar to the right of the figure.

The phase of the periodic branch formed by the two stable parts and unstable part (not indicated) traverses a large portion of the 0-180 degree phase envelope as the excitation frequency is increased over the range illustrated. Three sub-plots are also provided, illustrating snapshots of the beam deflection at three sample locations along the upper branch labelled A--C. In the 8Hz and 10Hz sub-plots the oscillatory response of the flexible beam is observed to correlate closely to the second linear mode shape of the system with a clearly defined nodal point observable. At the 9.64 Hz sample location this vibration mode and characteristic nodal point is not cleanly observed. Looking to the upper branch topology at this sample location one notes a localised `kink' in the periodic curve and accompanying variation in phase lag. This feature corresponds to a 3:1 resonance between the 2nd and 3rd nonlinear flexible modes (for reference, the first three linear modal frequencies for the vertically oriented beam are 1.44Hz, 10.42Hz and 29.55Hz). Note that the precise topology of the periodic structure around this resonance point is not known as the illustrated curve simply reflects the transient response as the base excitation sweep traverses this region; however, clearly the current formulation is capable of capturing these higher order resonances. For the further discussion of such resonant branch phenomena see \cite{Pai1990} wherein the out-of-plane resonance of a similar square cross-sectioned flexible beam is detailed.

Following the upper periodic curve towards the left, one notes the phase lag approaches (and passes through) 90 degrees. Nonlinear vibration theory dictates that (providing the strain rate damping law is linear) this point of quadrature at 90 degrees coincides with a point on the nonlinear normal mode (or backbone curve) of the flexible system. By allowing the unforced system to decay from this point of quadrature one may trace a portion of this backbone curve by sampling the transient frequency and amplitude of the decaying response. The resulting backbone section is shown in the figure by the thinner red curve, beginning at the quadrature point and decaying in amplitude until the linear frequency at 10.42Hz is met (note that the exact backbone curve for this system extends indefinitely beyond this quadrature point). The overall curve has a clear left tilt in the figure and thus indicates a softening of the second nonlinear normal mode as the amplitude of oscillation is increased (also observed in \cite{Pai2007}). One notes that the resonant `kink' previously observed in the upper branch also helps shape the transient decay and subsequent backbone in the region 9.5--10 Hz.

As in the previous examples validation is sought via comparison with the intrinsic beam formulation; performing an identical decay from the same initial state yields a second backbone approximation predicted by the latter method and indicated by the grey curve. One notes close agreement between each backbone result, both tracing a similar shape and indicating the nonlinear softening effect. Note that there is some slight difference between the two results in the 9.5--10 Hz range; however, due to the speed of decay it is unlikely that either method closely captures the localised backbone structure in this region. For the precise tracking of this backbone curve one could employ more sophisticated methods based upon shooting or pseudo-arclength continuation \cite{Nayfeh2007,Hill2017,Saghafi2015}, however, this is not pursued here.


\subsection{Wind Turbine Blade}\label{sec: Wind Turbine Blade}

\begin{figure}[t]
\centering
\includegraphics[width=0.90\textwidth,clip=true,trim= 94mm 20mm 84mm 32mm]{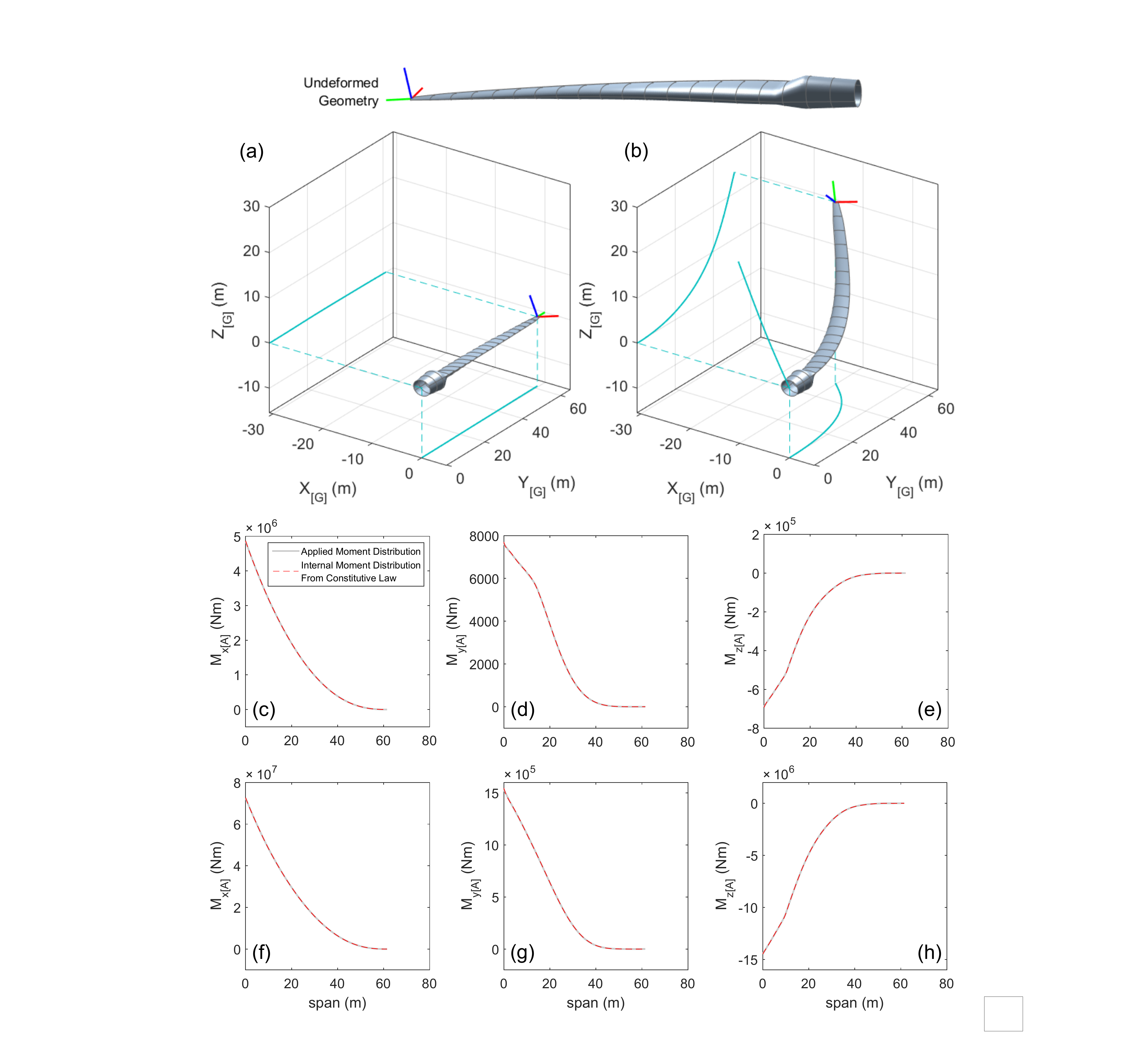}
\caption{Example deflections for the wind turbine test case.}
\label{fig: wind_turbine_multi_plot}
\end{figure}

The final test case details the static deformation of a wind turbine blade based upon the ``NREL offshore 5-MW baseline wind turbine'' detailed in the technical report of Jonkman \etal \cite{Jonkman2009} with supplementary parameters taken from the pre-design study of Kooijman \etal \cite{Kooijman2003} (part of the Dutch Offshore Wind Energy Converter (DOWEC) project). The blade has a span of 61.5m from it's hub attachment point to the tip; the unloaded structural geometry exhibits variable material properties, pre curvature and twist from its root to tip. For the specific turbine blade of this test, arbitrary parameterised functions are fitted to the data and used to concisely express these span dependent quantities; these functions are detailed in table \ref{table: turbine_blade_parameters}. A depiction of the undeformed blade geometry is provided at the top of figure \ref{fig: wind_turbine_multi_plot}.
\begin{table}[h]
\vspace{-0.3cm}
\centering
\begin{tabular}{|l|c|}
  \hline
  Span Dependent Function & Coefficients \\
  \hline
  $\msz{\mathrm{Length:\;\;}L = 61.5}{\hs{32}}$ $\mathrm{(m)}$ & \\ \hline
  $\hs{-2}\begin{array}{l}
      EI_{xx} = a_1+a_2x^{1/a_3} + a_4x + a_5x^2 + a_6x^3 + a_7x^4 \\
      \msz{(\mathrm{Nm^2})}{\hs{30}}x = 1-s/L
      \end{array}$ &
      $\begin{matrix}
      a_1 = 12.04  & a_2 = 7.4296 \\ a_3 = 5.9294 & a_4 = 0.6791 \\
      a_5 = 6.3115 & a_6 = -2.6386 \\ a_7 = -0.9242
      \end{matrix}$ \\ \hline
  $\hs{-2}\begin{array}{l}
      GJ = a_1+a_2x^{1/a_3} + a_4x + a_5x^2 + a_6x^3 + a_7x^4 \\
      \msz{(\mathrm{Nm/rad})}{\hs{30}}x = 1-s/L
      \end{array}$ &
      $\begin{matrix}
      a_1 = 12.15  & a_2 = 6.8864 \\ a_3 = 3.5466 & a_4 = -1.2809 \\
      a_5 = 2.0892 & a_6 = 2.8858 \\ a_7 = 0.0322
      \end{matrix}$ \\ \hline
  $\hs{-2}\begin{array}{l}
      EI_{zz} = a_1+a_2x^{1/a_3} + a_4x + a_5x^2 + a_6x^3 + a_7x^4 \\
      \msz{(\mathrm{Nm^2})}{\hs{30}}x = 1-s/L
      \end{array}$ &
      $\begin{matrix}
      a_1 = 15.43  & a_2 = 7.2698 \\ a_3 = 4.0351 & a_4 = 0.1553 \\
      a_5 = -0.2459 & a_6 = 0.3072 \\ a_7 = 0.4667
      \end{matrix}$ \\ \hline
    $\hs{-2}\begin{array}{l}
      m = a_1x^{1/2}\left(1 + a_2x^2 + a_3x^3 + a_4x^4 + a_5x^5 + a_6x^6\right) \\
      \msz{(\mathrm{kg/m})}{\hs{30}}x = 1-s/(L+0.2)
      \end{array}$ &
      $\begin{matrix}
      a_1 = 250 & a_2 = -1.4119 \\ a_3 = 17.8617 & a_4 = -21.5880 \\
      a_5 = -9.7330 & a_6 = 17.0449
      \end{matrix}$ \\ \hline
    $\hs{-2}\begin{array}{l}
      \theta_0 = -\tan^{-1}\left(\left(\sum\limits_{n=0}^{6}a_nx^n\right)(\tan^{-1}(s-14)/\pi+0.5)\right) \\
      \msz{(\mathrm{rad})}{\hs{30}}x = (s-30.75)/18.02
      \end{array}$ &
      $\hs{-1}\begin{matrix}
      a_0 = 0.029954 & a_1 = 0.039748 \\ a_2 = -0.0031189 & a_3 = -0.0074728 \\
      a_4 = 0.0080149 & a_5 = 0.00047782 \\ a_6 = -0.0021779 &
      \end{matrix}\hs{-1}$ \\ \hline
    $\msz{\psi_0 = 0}{\hs{32}}\mathrm{(rad)}$ & --- \\ \hline
    $\hs{-2}\begin{array}{l}
      \phi_0 = \left(13.5 - \sum\limits_{n=0}^{9}a_nx^n\right)(\tan^{-1}(s-10)/\pi+0.5)-13.5 \\
      \msz{(\mathrm{rad})}{\hs{30}}x = (s-30.75)/18.02
      \end{array}\hs{-4}$ &
      $\begin{matrix}
      a_0 = 6.5963 & a_1 = -5.3307 \\ a_2 = -0.71825 & a_3 = 0.73322 \\
      a_4 = 2.6739 & a_5 = -1.2031 \\ a_6 = -1.7149 & a_7 = 0.80577 \\
      a_8 = 0.30529 & a_9 = -0.14539
      \end{matrix}$ \\ \hline
\end{tabular}
\caption{Wind Turbine Blade Parameters.}
\vspace{-0.5cm}
\label{table: turbine_blade_parameters}
\end{table}
For this test the turbine blade is oriented such that
\begin{equation}
  \R{G}{A} = \mb
      \cos(20) & 0 & -\sin(20) \\
      0        & 1 & 0 \\
      \sin(20) & 0 & \cos(20) \me\,.
\end{equation}
$\svec{R}{E}$ is equal to the identity matrix. A distributed load is applied to this structure along the reference line $\svec{\Gamma}{G}(s)$. The load has an elliptical profile and acts normal to the blade chord line, i.e.
\begin{equation}
  \svec{F}{G}(s) = \bar{f}\sqrt{(1-(s/L)^2)}\;\svec{e}{z}(s)\;.
  \label{eq: wind_turbine_distributed_load}
\end{equation}
Two load cases corresponding to values of $\bar{f} = (2.4\e{5}/L)\,\mathrm{Nm^{-1}}$ and $\bar{f} = (4\e{6}/L)\,\mathrm{Nm^{-1}}$ in expression (\ref{eq: wind_turbine_distributed_load}) are applied to the blade in panels (a) and (b) of figure \ref{fig: wind_turbine_multi_plot}, respectively. Each panel shows the static deformation achieved via dynamic simulation of the system, allowing sufficient time for transient motion to settle; a conservative choice of eight Chebyshev polynomials per Euler angle was used for the treatment of this test. For the first load case of panel (a) the applied load is just sufficient to remove the pre-curvature from the turbine blade. For the second load case of panel (b) the load produces an extreme deflection of the structure far in excess of any physically realisable load; this latter case is used to demonstrate the ability to solve the geometrically exact problem for large deformations of the generic non-prismatic structure. To verify these deformations both the net applied moment (a consequence of the applied load distribution $\svec{F}{G}$) and the internal moment (calculated from the Euler Bernoulli constituent relation) are compared at each spanwise location in the range $s\in[0,L]$. In panel (c) the out-of-plane ($\svec{e}{x}$) component of the external (solid) and internal (dashed) moment distributions are plotted for the first load case; panels (d) and (e) depict the twist ($\svec{e}{y}$) and in-plane ($\svec{e}{z}$) components respectively. Panels (f)--(h) provide the same three plots for the second, larger load case. Correct solution of the static problem may be verified by noting the close agreement indicated in these sub plots.

\section{Conclusions}\label{sec: Conclusions}
This paper has detailed the development and application of a \methodName approach capable of the low order geometrically exact representation of a flexible beam; the formulation is based upon the shape based discretisation of attitude and incremental shear kinematic quantities distributed along the flexible structure. The development of the describing equations of motion was demonstrated for an example Euler angle parameterisation and coupled to a set of Chebyshev polynomials of the first kind, modified to admit the generalised kinematic condition. The formulation was subsequently applied to a variety of test cases including the modelling of planar and non-planar transient dynamics, the prediction of static equilibria, the treatment of simple beam, pre-curved and non-prismatic structures, all within a geometrically nonlinear context. The shape functions used are generic and may be tailored to the characteristics of the modelled physical system. Complementary Nastran and intrinsic beam calculations were used to verify specific results throughout these tests; altogether the formulation of this study demonstrated the accurate treatment of the considered test cases, consistently achieved using a small set of states, representing an order of magnitude reduction in problem size over its element-based counterparts.
This low order formulation admits the efficient treatment of flexible beam analyses for which the size of system generated by traditional element based representations would incur a significant computational penalty; such examples include structural optimisation problems over large parametric design spaces, sensitivity and uncertainty quantification analyses, and numerical continuation methods exploring limiting dynamic behaviours.

\ethics{
Not applicable to this work.}

\dataccess{
Datasets supporting the depicted results of this study will be made available online.}

\conflict{
The authors declare that they have no competing interests.}

\authorCntrb{
Chris Howcroft led the development of the \methodName formulation and numerical testing campaign; Robbie Cook supported the study by constructing the intrinsic beam element code from which validating results were drawn. Simon Neild and Mark Lowenberg helped identify the proposed test cases and all authors contributed to the preparation of the manuscript and gave final approval for publication.}

\funding{
The research leading to these results has received funding from the InnovateUK Agile Wing Integration Project (TSB-113041) and the AEROGUST project funded from the European Union's Horizon 2020 research and innovation programme (grant No. 636053). Simon Neild is supported by an EPSRC fellowship (EP/K005375/1) and Jonathan Cooper holds a Royal Academy of Engineering Research Chair.}

\ack{
We gratefully acknowledge the support of our funders.}


\appendix
\section{Appendices}\label{Asec: appendix}

\subsection{Virtual Work Derivatives}\label{Asec: work derivatives}

{\bf Strain:} From equation (\ref{eq:delta W_K}),
\begin{flalign}
    \pdiff{W_K}{q_{jk}} = \intt{0}{L}{F}_{K}(s,\vec{\Delta\xi})_i \; \pdiff{\xi_i}{q_{jk}} \d s
    \label{eq:pdiff{W_K}{q}}
  \;,\hs{5}\mathrm{where}\hs{5}
  \pdiff{\xi_i}{q_{jk}} = \pdiff{\xi_i}{\zeta_j}\pdiff{\zeta_j}{q_{jk}}
                    + \pdiff{\xi_i}{\zeta_j^\prime}\pdiff{\zeta_j^\prime}{q_{jk}}\;,
\end{flalign}
$F_K(s,\vec{\Delta\xi})_i$ denoting the $ith$ component of $\vec{F_K}$.
\\ \\
{\bf Damping:} From equation (\ref{eq:delta W_C}),
\begin{flalign}
    \pdiff{W_C}{q_{jk}} = \intt{0}{L}{F}_{C}(s,\vec{\Delta\xi},\vec{\dot\xi})_i \; \pdiff{\xi_i}{q_{jk}} \d s
    \label{eq:pdiff{W_C}{q}}
    \;,\hs{5}\mathrm{where}\hs{5}
    \dot\xi_i = \left(\pdiff{\xi_i}{\zeta_j}\pdiff{\zeta_j}{t}
                    + \pdiff{\xi_i}{\zeta_j^\prime}\pdiff{\zeta_j^\prime}{t}\right)\;.
\end{flalign}
\\ \\
{\bf Applied Force:} From equation (\ref{eq:delta W_F}),
\begin{flalign}
    \pdiff{W_F}{q_{jk}} = \intt{0}{L} \svec{F}{[G]}(s) \cdot \pdiff{\svec{r}{F[G]}}{q_{jk}} \d s
    \;,\hs{5}\mathrm{where}\hs{5}
    \pdiff{\svec{r}{F[G]}}{q_{jk}} &= \;\pdiff{\svec{\E}{[G]}}{q_{jk}}\mb a_F \\ 0 \\ c_F \me + \pdiff{\svec{\Gamma}{[G]}}{q_{jk}}\;.
    \label{eq:pdiff{W_F}{q}}
\end{flalign}
\\ \\
{\bf Applied Moment:} From equation (\ref{eq:delta W_M}),
\begin{flalign}
    \pdiff{W_M}{q_{jk}} = \intt{0}{L} \svec{M}{[G]}(s) \cdot \pdiff{\svec{\vartheta}{[G]}}{q_{jk}} \d s
    \label{eq:pdiff{W_M}{q}}\;.
\end{flalign}
\\ \\
{\bf Kinetic Term:} From equation (\ref{eq:delta W_T}),
\begin{flalign}
  && &\pdiff{W_T}{q_{jk}} = \intt{0}{L} m\svec{\ddot r}{m[G]}\cdot\pdiff{\svec{r}{m[G]}}{q_{jk}}
             + \left(\svec{\dot\E}{[G]}\svec{I}{\vartheta[I]}\vec{\E_{[G]}^T}\svec{\dot\vartheta}{[G]}
    + \svec{\E}{[G]}\svec{I}{\vartheta[I]}\vec{\E_{[G]}^T}\svec{\ddot\vartheta}{[G]}\right)\cdot\pdiff{\svec{\vartheta}{[G]}}{q_{jk}} \;\d s\;,
             \label{eq:pdiff{W_T}{q}}
\end{flalign}
\begin{flalign*}
  \pdiff{\svec{r}{m[G]}}{q_{jk}} &= \;\pdiff{\svec{\E}{[G]}}{q_{jk}}\mb a_m \\ 0 \\ c_m \me + \pdiff{\svec{\Gamma}{[G]}}{q_{jk}}
  \;,\hs{5}\mathrm{where}\hs{5}
  \svec{\ddot r}{m[G]} \;=\; \svec{\ddot\E}{[G]}\mb a_m \\ 0 \\ c_m \me + \svec{\ddot\Gamma}{[G]}\;.
\end{flalign*}
\\ \\
To complete the terms (\ref{eq:pdiff{W_K}{q}})--(\ref{eq:pdiff{W_T}{q}}) one requires definition of the partial derivatives
${\partial \svec{\Gamma}{[G]}}/{\partial q_{jk}}$\:,\:
${\partial^2\svec{\Gamma}{[G]}}/{\partial t^2}$\:,\:
${\partial \svec{\vartheta}{[G]}}/{\partial q_{jk}}$\:,\:
${\partial^2\svec{\vartheta}{[G]}}/{\partial t^2}$\:,\:
${\partial \svec{\E}{[G]}}/{\partial q_{jk}}$\:,\:
${\partial \svec{\E}{[G]}}/{\partial t}$\:,\:
${\partial^2 \vec\E}/{\partial t^2}$\:,\:
${\partial{\xi_i}/\partial\zeta_j}$\:,\:
${\partial{\xi_i}/\partial\zeta_j^\prime}$\:,\:
$\partial\zeta_j/\partial q_{jk}$\:,\:
$\partial\zeta_j^\prime/\partial q_{jk}$\:,\:
$\partial\zeta_j/\partial t$\:,\:
$\partial\zeta_j^\prime/\partial t$\:,\:
$\partial^2\zeta_j/\partial t^2$\:.\:
These are given in the following sections \nolink{\ref{Asec: appendix}}\ref{Asec: zeta derivatives}--\nolink{\ref{Asec: appendix}}\ref{Asec: Gamma derivatives}.

\subsection{\texorpdfstring{$\vec{\zeta}$ Derivatives}{\textzeta \; Derivatives}}\label{Asec: zeta derivatives}

\begin{flalign}
    &\zeta_j(s,t) = B_{jk}(s)\,q_{jk}(t)+\zeta_{j0}(s)\notag\\[2mm]
    &\pdiff{\zeta_j}{q_{jk}} = B_{jk}\hs{2},\hs{3}
    \pdiff{\zeta_j^\prime}{q_{jk}} = B^\prime_{jk}\hs{2},\hs{3}
    \pdiff{\zeta_j}{t} = B_{jk}\,\dot q_{jk}\hs{2},\hs{3}
    \pdiff{\zeta_j^\prime}{t} = B^\prime_{jk}\,\dot q_{jk}\hs{2},\hs{3}
    \pdiff{^2\zeta_j}{t^2} = B_{jk}\,\ddot q_{jk}
    \label{eq:zeta shape series}
\end{flalign}

\subsection{\texorpdfstring{$\vec{\xi}$ Derivatives}{\textxi \; Derivatives}}\label{Asec: xi derivatives}

The partial derivatives ${\partial{\xi_i}/\partial\zeta_j}$ and ${\partial{\xi_i}/\partial\zeta_j^\prime}$ follow simply from differentiation of
\begin{flalign}
  \vec{\xi} = \mb \tau_x \\ \varepsilon \\ \tau_z \\ \kappa_x \\ \kappa_y \\ \kappa_z \me
      = \mb \tau_x \\ \varepsilon \\ \tau_z \\
            \svec{R}{E}
            \begin{bmatrix}\psi^\prime\ct\sp+\theta^\prime\cp \\
            \phi^\prime-\psi^\prime\st \\
           -\psi^\prime\ct\cp+\theta^\prime\sp
            \end{bmatrix} \me
             \mathrm{with\,respect\,to} \hs{1}
  \vec{\zeta} = \mb \tau_x \\ \varepsilon \\ \tau_z \\ \theta \\ \psi \\ \phi \me
             \mathrm{and} \hs{1}
  \vec{\zeta^\prime} = \mb \tau_x^\prime \\ \varepsilon^\prime \\ \tau_z^\prime \\ \theta^\prime \\ \psi^\prime \\ \phi^\prime \me
\end{flalign}

\subsection{\texorpdfstring{$\vec{\vartheta}$ Derivatives}{Rotational Derivatives}}\label{Asec: vartheta derivatives}

$\svec{\delta\vartheta}{[A]}(\vec{\delta\zeta})$ is given by equation (\ref{eq:delta vartheta}). Thus each partial derivative of $\svec{\vartheta}{[A]}$ follows from differentiation of $\vec{\delta\zeta}$ with respect to the variable of interest (see \nolink{\ref{Asec: appendix}}\ref{Asec: zeta derivatives}).
\\ \\
$\svec{\ddot\vartheta}{[A]}(\vec{\ddot\zeta})$ is given by equation (\ref{eq:ddot vartheta}). Similarly $\vec{\ddot\zeta}$ is given in \nolink{\ref{Asec: appendix}}\ref{Asec: zeta derivatives}.
\\ \\
In the global coordinate system
\begin{flalign}
  && &\svec{\delta\vartheta}{[G]} = \svec{\delta\beta}{A[G]} + \R{G}{A}\svec{\delta\vartheta}{[A]}
  \hs{1},\hs{1}
  \svec{\ddot\vartheta}{[G]} = \svec{\dot{\bar\Omega}}{A[G]} + \R{G}{A}\svec{\ddot\vartheta}{[A]} + \vec{\diff{\R{G}{A}}{t}}\svec{\dot\vartheta}{[A]} \\
  &\mathrm{where}& &\vec{\diff{\R{G}{A}}{t}} \;=\; \svec{\bar\Omega}{A[G]}\times\R{G}{A}\;. \notag
\end{flalign}
$\svec{\beta}{A}$ is the rotation vector of the reference frame ${\bf A}$ and $\svec{\bar\Omega}{A} = \svec{\dot\beta}{A}$.

\subsection{\texorpdfstring{$\vec{\Gamma}$  Derivatives}{Translational Derivatives}}\label{Asec: Gamma derivatives}

\begin{flalign}
  \svec{\delta\E}{[G]} = \svec{\delta\vartheta}{[G]}\times\svec{\E}{[G]}
  \hs{2},\hs{4}
  \svec{\ddot\E}{[G]} = \svec{\ddot\vartheta}{[G]}\times\svec{\E}{[G]} + \svec{\dot\vartheta}{[G]}\times\svec{\dot\vartheta}{[G]}\times\svec{\E}{[G]}
\end{flalign}
\begin{flalign}
  &\svec{\Gamma}{[G]} = \svec{\bar r}{A[G]} + \intt{0}{s}
  \svec{\E}{[G]}
  \mb\tau_x\\1+\varepsilon\\\tau_z\me
  \d \tilde s
  \hs{1} = \hs{1} \svec{\bar r}{A[G]} + \intt{0}{s} \svec{\E}{[G]} {\Scale[1.6]{{\tau}}} \d \tilde s && \\
  &\svec{\delta\Gamma}{[G]}
  = \svec{\dot{\bar r}}{A[G]} + \intt{0}{s} \svec{\delta\E}{[G]} {\Scale[1.6]{{\tau}}} + \svec{\E}{[G]} \vec\delta{\Scale[1.6]{{\tau}}} \d \tilde s\hs{1},\hs{2}
  \svec{\ddot\Gamma}{[G]}
  = \svec{\ddot{\bar r}}{A[G]} + \intt{0}{s} \svec{\ddot\E}{[G]} {\Scale[1.6]{{\tau}}} + 2\svec{\dot\E}{[G]} {\Scale[1.6]{{\dot\tau}}} + \svec{\E}{[G]} {\Scale[1.6]{{\ddot\tau}}} \d \tilde s\;. &&
\end{flalign}

\subsection{Re-arrangement of Kinetic Terms}\label{Asec: Re-arrangement of Kinetic Terms}
Rearrangement of the 2nd order derivative terms into the form $(a\vec{\ddot q}+b)$ is required to cast this formulation as the explicit system of ODEs (\ref{eq:EoM}). These rearrangements are detailed here.

\begin{flalign*}
  \svec{\delta\vartheta}{[I]}(\vec{\delta\zeta}) &= \delta\theta\mb\cs\\-\ss\\0\me + \delta\psi\mb0\\0\\-1\me
    + \delta\phi \: \svec{y}{E} & \\[4mm]
  \svec{\dot\vartheta}{[I]}(\vec{\zeta},\vec{e_\chi}) &= \pdiff{\vartheta}{\zeta_j}\pdiff{\zeta_j}{q_{jk}}\pdiff{q_{jk}}{t}
    = \dot\theta\mb\cs\\-\ss\\0\me + \dot\psi\mb0\\0\\-1\me+\dot\phi \: \svec{y}{E} &
\end{flalign*}
\begin{flalign*}
  \svec{\ddot\vartheta}{[I]}(\vec{\zeta},\vec{e_\chi})
    &= \pdiff{\vartheta}{\zeta_j}\pdiff{\zeta_j}{q_{jk}}\pdiff{^2q_{jk}}{t^2}
    + \pdiff{}{t}\left(\pdiff{\vartheta}{\zeta_j}\pdiff{\zeta_j}{q_{jk}}\right)\pdiff{q_{jk}}{t} &\notag\\
    &= \pdiff{\vartheta}{\zeta_j}\pdiff{\zeta_j}{q_{jk}}\pdiff{^2q_{jk}}{t^2}
    + \pdiff{}{\zeta_j}\left(\pdiff{\vartheta}{\zeta_j}\pdiff{\zeta_j}{q_{jk}}\right)\pdiff{\zeta_j}{t}\pdiff{q_{jk}}{t}
    + \pdiff{}{e_\chi}\left(\pdiff{\vartheta}{\zeta_j}\pdiff{\zeta_j}{q_{jk}}\right)\pdiff{e_\chi}{t}\pdiff{q_{jk}}{t} &\notag\\
    &= \left[\pdiff{\vartheta}{q_{jk}}\pdiff{^2q_{jk}}{t^2}\right]
    + \left[\pdiff{\zeta_i}{t}\left(\frac{\partial^2\vartheta}{\partial \zeta_i\partial\zeta_j}\right)\pdiff{\zeta_j}{t}\right]
    + \left[\pdiff{e_\chi}{t}\left(\frac{\partial^2\vartheta}{\partial e_\chi\partial\zeta_j}\right)\pdiff{\zeta_j}{t}\right] &\notag\\
    &=
    \left[\ddot\theta\mb\cs\\-\ss\\0\me + \ddot\psi\mb0\\0\\-1\me + \ddot\phi \: \svec{y}{E}\right]
  + \left[\dot\theta\dot\psi\mb-\ss\\-\cs\\0\me\right] + \left[\dot\phi \: \svec{\dot y}{E}\right] & \\
\end{flalign*}
\begin{flalign}
  &\svec{\ddot\vartheta}{[I]}
    = \left(\vec{\pdiff{\vartheta_{[I]}}{q_{jk}}}\right)\pdiff{^2q_{jk}}{t^2} + \vec{\ddot\vartheta_{[I]}^*} && \notag\\
  &\mathrm{where}\hs{5} \vec{\ddot\vartheta_{[I]}^*} = \dot\theta\dot\psi\mb-\ss\\-\cs\\0\me + \dot\phi \: \svec{\dot y}{E}\;. &&
\end{flalign}

\begin{flalign}
  &\svec{\ddot\vartheta}{[G]}
    = \left(\vec{\pdiff{\vartheta_{[G]}}{q_{jk}}}\right)\pdiff{^2q_{jk}}{t^2} + \vec{\ddot\vartheta_{[G]}^*} && \notag\\
  &\mathrm{where}\hs{5} \vec{\ddot\vartheta_{[G]}^*} = \vec{{\dot{\bar\Omega}}_{A[G]}^*}
      + \vec{{\bar\Omega}_{A[G]}}\times\svec{\dot\vartheta}{[G]} + \R{G}{A}\svec{R}{E}
      \svec{\ddot\vartheta}{[I]}^*\;. &&
\end{flalign}
Note if $\svec{\bar\Omega}{A[G]}$ is prescribed (i.e. not state dependent) then $\vec{\dot{\bar\Omega}_{A[G]}^*} = \svec{\dot{\bar\Omega}}{A[G]}$.

\begin{flalign}
  &\svec{\ddot\E}{[G]}
    = \left(\vec{\pdiff{\E_{[G]}}{q_{jk}}}\right)\pdiff{^2q_{jk}}{t^2} + \vec{\ddot\E_{[G]}^*} && \notag\\
  &\mathrm{where}\hs{5} \vec{\ddot\E_{[G]}^*} = \vec{\ddot\vartheta_{[G]}^*}\times\svec{\E}{[G]} + \svec{\dot\vartheta}{[G]}\times\svec{\dot\vartheta}{[G]}\times\svec{\E}{[G]}\;. &&
\end{flalign}

\begin{flalign}
  &\svec{\ddot r}{m[G]}
    = \left(\vec{\pdiff{r_{m[G]}}{q_{jk}}}\right)\pdiff{^2q_{jk}}{t^2} + \vec{\ddot r_{m[G]}^*} && \notag\\
  &\mathrm{where}\hs{5} \vec{\ddot r_{m[G]}^*} \;=\; \vec{{\ddot{\bar r}}_{A[G]}^*} + \vec{{\ddot\E}_{[G]}^*}\mb a_m \\ 0 \\ c_m \me
    + \intt{0}{s} \vec{{\ddot\E}_{[G]}^*} {\Scale[1.6]{{\tau}}} + 2\svec{\dot\E}{[G]} {\Scale[1.6]{{\dot\tau}}} + \svec{\E}{[G]} \cancelto{0}{\Scale[1.6]{{\ddot\tau}}^{\vec *}\;} \d \tilde s\;. &&
  \end{flalign}
If $\svec{\bar r}{A[G]}$ is prescribed then $\vec{{\ddot{\bar r}}_{A[G]}^*} = \vec{\ddot{\bar r}}_{A[G]}$.

\bibliographystyle{unsrt}
\bibliography{bibliography}

\begin{thebibliography}{46}
\providecommand{\natexlab}[1]{#1}
\providecommand{\url}[1]{\texttt{#1}}
\expandafter\ifx\csname urlstyle\endcsname\relax
  \providecommand{\doi}[1]{doi: #1}\else
  \providecommand{\doi}{doi: \begingroup \urlstyle{rm}\Url}\fi

\bibitem[Reissner(1973)]{Reissner1973}
Reissner E.
\newblock {On One-Dimensional Large-Displacement Finite-Strain Beam Theory}.
\newblock \emph{Studies in Applied Mathematics}, 52\penalty0 (2):\penalty0
  87--95, 1973.

\bibitem[Simo and Vu-Quoc(1985)]{Simo_Pt1}
Simo J.~C and Vu-Quoc L.
\newblock {A Finite Strain Beam Formulation. The Three-Dimensional Dynamic
  Problem. Part I.}
\newblock \emph{Computer Methods in Applied Mechanics and Engineering},
  49:\penalty0 55--70, 1985.

\bibitem[Simo and Vu-Quoc(1986)]{Simo_Pt2}
Simo J.~C and Vu-Quoc L.
\newblock {A Three-Dimensional Finite-Strain Rod Model. Part II: Computational
  Aspects}.
\newblock \emph{Computer Methods in Applied Mechanics and Engineering},
  58:\penalty0 76--116, 1986.

\bibitem[Bauchau et~al.(2014)Bauchau, Han, and Mikkola]{Bauchau2014}
Bauchau O.~A, Han S, and Mikkola A.
\newblock {Comparison of the Absolute Nodal Coordinate and Geometrically Exact
  Formulations for Beams}.
\newblock \emph{Multibody System Dynamics}, 32:\penalty0 67--85, 2014.

\bibitem[Shabana et~al.(1998)Shabana, Hussien, and Escalona]{Shabana1998}
Shabana A.~A, Hussien H.~A, and Escalona J.~L.
\newblock {Application of the Absolute Nodal Coordinate Formulation to Large
  Rotation and Large Deformation Problems}.
\newblock \emph{Journal of Mechanical Design}, 120\penalty0 (2):\penalty0
  188--195, 1998.

\bibitem[Wempner(1969)]{Wempner1969}
Wempner G.
\newblock {Finite Elements, Finite Rotations and Small Strains of Flexible
  Shells}.
\newblock \emph{International Journal of Solids and Structures}, 5\penalty0
  (2):\penalty0 117--153, 1969.

\bibitem[Belytschko and Hsieh(1973)]{Belytschko1973}
Belytschko T and Hsieh B.~J.
\newblock {Nonlinear Transient Finite Element Analysis with Convected
  Coordinates}.
\newblock \emph{International Journal for Numerical Methods in Engineering},
  7\penalty0 (3):\penalty0 255--271, 1973.

\bibitem[Belytschko and Glaum(1979)]{Belytschko1979}
Belytschko T and Glaum L.~W.
\newblock {Applications of Higher Order Corotational Stretch Theories to
  Nonlinear Finite Element Analysis}.
\newblock \emph{Computers \& Structures}, 10\penalty0 (1):\penalty0 175--182,
  1979.

\bibitem[Cesnik and Brown(2002)]{Cesnik2002}
Cesnik C.~E.~S and Brown E.~L.
\newblock {Modeling of High Aspect Ratio Active Flexible Wings for Roll
  Control}.
\newblock In \emph{43rd AIAA/ASME/ASCE/AHS/ASC Structures, Structural Dynamics,
  and Materials Conference}, Denver, Colorado, 2002.

\bibitem[Santos et~al.(2011)Santos, Pimenta, and Almeida]{Santos2011_etal}
Santos H.~A. F.~A, Pimenta P.~M, and Almeida J.~P.~M.
\newblock {A Hybrid-mixed Finite Element Formulation for the Geometrically
  Exact Analysis of Three-dimensional Framed Structures}.
\newblock \emph{Computational Mechanics}, 48:\penalty0 591--613, 2011.

\bibitem[Prathap and Bhashyam(1982)]{Prathap1982}
Prathap G and Bhashyam G.~R.
\newblock {Reduced Integration and the Shear-Flexible Beam Element}.
\newblock \emph{International Journal for Numerical Methods in Engineering},
  18:\penalty0 195--210, 1982.

\bibitem[Santos(2011)]{Santos2011_overview}
Santos H.~A. F.~A.
\newblock {Complementary-Energy Methods for Geometrically Non-linear Structural
  Models: An Overview and Recent Developments in the Analysis of Frames}.
\newblock \emph{Archives of Computational Methods in Engineering}, 18:\penalty0
  405--440, 2011.

\bibitem[Hodges(1990)]{Hodges1990}
Hodges D.~H.
\newblock {A Mixed Variational Formulation Based on Exact Intrinsic Equations
  or Dynamics of Moving Beams}.
\newblock \emph{International Journal of Solids and Structures}, 26\penalty0
  (11):\penalty0 1253--1273, 1990.

\bibitem[Patil et~al.(2001)Patil, Hodges, and Cesnik]{Patil2001}
Patil M.~J, Hodges D.~H, and Cesnik C.~E.~S.
\newblock {Nonlinear Aeroelasticity and Flight Dynamics of High-Altitude
  Long-Endurance Aircraft}.
\newblock \emph{Journal of Aircraft}, 38\penalty0 (1):\penalty0 88--94, 2001.

\bibitem[Wasfy and Noor(2003)]{Wasfy2003}
Wasfy T.~M and Noor A.~K.
\newblock {Computational Strategies for Flexible Multibody Systems}.
\newblock \emph{Applied Mechanics Reviews}, 56\penalty0 (6):\penalty0 553--613,
  2003.

\bibitem[Shabana(1997)]{Shabana1997}
Shabana A.~A.
\newblock {Flexible Multibody Dynamics: Review of Past and Recent
  Developments}.
\newblock \emph{Multibody System Dynamics}, 1\penalty0 (2):\penalty0 189--222,
  1997.

\bibitem[Hollkamp and Gordon(2008)]{Hollkamp2008}
Hollkamp J.~J and Gordon R.~W.
\newblock {Reduced-Order Models for Nonlinear Response Prediction: Implicit
  Condensation and Expansion}.
\newblock \emph{Journal of Sound and Vibration}, 318:\penalty0 1139--1153,
  2008.

\bibitem[Kuether and Allen(2015)]{Kuether2016}
Kuether R.~J and Allen M.~S.
\newblock {Validation of Nonlinear Reduced Order Models with Time Integration
  Targeted at Nonlinear Normal Modes}.
\newblock In \emph{Proceedings of the 33rd IMAC, A Congerence and Exposition on
  Structural Dynamics}, pages 363--375, Orlando, Florida, 2015.

\bibitem[Przekop et~al.(2012)Przekop, Guo, and Rizzi]{Przekop2012}
Przekop A, Guo X, and Rizzi S.~A.
\newblock {Alternative Modal Basis Selection Procedures for Reduced-Order
  Nonlinear Random Response Simulation}.
\newblock \emph{Journal of Sound and Vibration}, 331:\penalty0 4005--4024,
  2012.

\bibitem[Mignolet et~al.(2013)Mignolet, Przekop, Rizzi, and
  Spottswood]{Mignolet2013}
Mignolet M.~P, Przekop A, Rizzi S.~A, and Spottswood S.~M.
\newblock {A Review of Indirect/Non-Intrusive Reduced Order Modeling of
  Nonlinear Geometric Structures}.
\newblock \emph{Journal of Sound and Vibration}, 332:\penalty0 24337--2460,
  2013.

\bibitem[Rao(2007)]{Rao2007}
Rao S.~S.
\newblock \emph{{Vibration of Continuous Systems}}.
\newblock John Wiley \& Sons, Inc., Hoboken, New Jersey, 2007.

\bibitem[Patil and Althoff(2010)]{Patil2010}
Patil M.~J and Althoff M.
\newblock {Energy-consistent, Galerkin approach for the nonlinear dynamics of
  beams using mixed, intrinsic equations}.
\newblock \emph{Journal of Vibration and Control}, 17\penalty0 (11):\penalty0
  1748--1758, 2010.

\bibitem[Hodges et~al.(1996)Hodges, Shang, and Cesnik]{Hodges1996}
Hodges D.~H, Shang X, and Cesnik C.~E.~S.
\newblock {Finite Element Solution of Nonlinear Intrinsic Equations for Curved
  Composite Beams}.
\newblock \emph{Journal of the American Helicopter Society}, 41\penalty0
  (4):\penalty0 313--321, 1996.

\bibitem[Palacios et~al.(2010)Palacios, Murua, and Cook]{Palacios2010}
Palacios R, Murua J, and Cook R.~G.
\newblock {Structural and Aerodynamic Models in Nonlinear Flight Dynamics of
  Very Flexible Aircraft}.
\newblock \emph{AIAA Journal}, 48\penalty0 (11):\penalty0 2648--2659, 2010.

\bibitem[Antman(1974)]{Antman1974}
Antman S.~S.
\newblock {Kirchhoff's Problem for Nonlinearly Elastic Rods}.
\newblock \emph{Quarterly of Applied Mathematics}, 32\penalty0 (3):\penalty0
  221--240, 1974.

\bibitem[Hodges and Dowell(1974)]{Hodges1974}
Hodges D.~H and Dowell E.~H.
\newblock {Nonlinear Equations of Motion for the Elastic Bending and Torsion of
  Twisted Nonuniform Rotor Blades}.
\newblock Technical Report NASA TN D-7818, National Aeronautics and Space
  Administration, December 1974.

\bibitem[Yu et~al.(2002)Yu, Hodges, Volovoi, and Cesnik]{Yu2002}
Yu~W, Hodges D.~H, Volovoi V, and Cesnik C.~E.~S.
\newblock {On Timoshenko-Like Modeling of Initially Curved and Twisted
  Composite Beams}.
\newblock \emph{International Journal of Solids and Structures}, 39:\penalty0
  5101--5121, 2002.

\bibitem[Shuster(1993)]{Shuster1993}
Shuster M.~D.
\newblock {A Survey of Attitude Representations}.
\newblock \emph{The Journal of the Astronautical Sciences}, 41\penalty0
  (4):\penalty0 439--517, 1993.

\bibitem[Hesse(2013)]{Hesse2013}
Hesse H.
\newblock {Consistent Aeroelastic Linearisation and Reduced-Order Modelling in
  the Dynamics of a Manoeuvring Flexible Aircraft}.
\newblock Master's thesis, Imperial College London, August 2013.

\bibitem[Crisfield(1990)]{Crisfield1990}
Crisfield M.~A.
\newblock {A Consistent Co-Rotational Formulation for Non-Linear,
  Three-Dimensional, Beam-Elements}.
\newblock \emph{Computer Methods in Applied Mechanics and Engineering},
  81:\penalty0 131--150, 1990.

\bibitem[Pai(2007)]{Pai2007}
Pai P.~F.
\newblock \emph{{Highly Flexible Stuctures: Modeling, Computation, and
  Experimentation}}.
\newblock American Institute of Aeronautics and Astronautics, Inc., Reston,
  Virginia, 2007.

\bibitem[Ibrahimbegovi\'c et~al.(2000)Ibrahimbegovi\'c, Momouri, Taylor, and
  Chen]{Ibrahimbegovic2000}
Ibrahimbegovi\'c A, Momouri S, Taylor R.~L, and Chen A.~J.
\newblock {Finite Element Method in Dynamics of Flexible Multibody Systems:
  Modeling of Holonomic Constraints and Energy Conserving Integration Schemes}.
\newblock \emph{Multibody System Dynamics}, 4:\penalty0 195--223, 2000.

\bibitem[Shampine and Reichelt(1997)]{Shampine1997}
Shampine L.~F and Reichelt M.~W.
\newblock {The Matlab ODE Suite}.
\newblock \emph{SIAM Journal on Scientific Computing (SISC)}, 18\penalty0
  (1):\penalty0 1--22, 1997.

\bibitem[Bathe and Bolourchi(1979)]{Bathe1979}
Bathe K and Bolourchi S.
\newblock {Large Displacement Analysis of Three-Dimensional Beam Structures}.
\newblock \emph{International Journal for Numerical Methods in Engineering},
  14:\penalty0 961--986, 1979.

\bibitem[G\'{e}radin and Cardona(2001)]{Geradin2001}
G\'{e}radin M and Cardona A.
\newblock \emph{{Flexible Multibody Dynamics: A Finite Element Approach}}.
\newblock John Wiley \& Sons, 2001.

\bibitem[Li and Vu-Quoc(2010)]{Li2010}
Li~Z.~X and Vu-Quoc L.
\newblock {A Mixed Co-Rotational 3D Beam Element Formulation for Arbitrarily
  Large Rotations}.
\newblock \emph{Advanced Steel Construction}, 6\penalty0 (2):\penalty0
  767--787, 2010.

\bibitem[Pi et~al.(2007)Pi, Bradford, and F.]{Pi2007}
Pi~Y.~L, Bradford M.~A, and F.~T.-L.
\newblock {Nonlinear Analysis and Buckling of Elastically Supported Circular
  Shallow Arches}.
\newblock \emph{International Journal of Solids and Structures}, 44:\penalty0
  2401--2425, 2007.

\bibitem[Berzeri and Shabana(2002)]{Berzeri2002}
Berzeri M and Shabana A.~A.
\newblock {Study of the Centrifugal Stiffening Effect Using the Finite Element
  Absolute Nodal Coordinate Formulation}.
\newblock \emph{Multibody System Dynamics}, 7:\penalty0 357--387, 2002.

\bibitem[Fung and Yau(1999)]{Fung1999}
Fung E.~H.~K and Yau D.~T.~W.
\newblock {Effects of Centrifugal Stiffening on the Vibration Frequencies of a
  Constrained Flexibe Arm}.
\newblock \emph{Journal of Sound and Vibration}, 224:\penalty0 809--841, 1999.

\bibitem[Yang et~al.(2004)Yang, Jiang, and Chen]{Yang2003}
Yang J.~B, Jiang L.~J, and Chen D.~C.~H.
\newblock {Dynamic Modelling and Control of a Rotating Euler-Bernoulli Beam}.
\newblock \emph{Journal of Sound and Vibration}, 274:\penalty0 863--875, 2004.

\bibitem[Pai and Nayfeh(1990)]{Pai1990}
Pai P.~F and Nayfeh A.~H.
\newblock {Non-Linear Non-Planar Oscillations of a Cantilever Beam under
  Lateral Base Excitations}.
\newblock \emph{International Journal of Non-Linear Mechanics}, 25\penalty0
  (5):\penalty0 455--474, 1990.

\bibitem[Nayfeh and Balachandran(2007)]{Nayfeh2007}
Nayfeh A.~H and Balachandran B.
\newblock \emph{{Applied Nonlinear Dynamics, Analytical, Computational and
  Experimental Methods}}.
\newblock Wiley-VCH Verlag GmbH, Weinheim, Germany, 2007.

\bibitem[Hill et~al.(2017)Hill, Cammarano, Neild, and Barton]{Hill2017}
Hill T.~L, Cammarano A, Neild S.~A, and Barton D.~A.~W.
\newblock {Identifying the Significance of Nonlinear Normal Modes}.
\newblock \emph{Proceedings of the Royal Society A: mathematical, physical and
  engineering sciences}, 473\penalty0 (2199), 2017.

\bibitem[Saghafi et~al.(2015)Saghafi, Dankowicz, and Lacarbonara]{Saghafi2015}
Saghafi M, Dankowicz H, and Lacarbonara W.
\newblock {Nonlinear Tuning of Microresonators for Dynamic Range Enhancement}.
\newblock \emph{Proceedings of the Royal Society A: mathematical, physical and
  engineering sciences}, 471\penalty0 (2179), 2015.

\bibitem[Jonkman et~al.(2009)Jonkman, Butterfield, Musial, and
  Scott]{Jonkman2009}
Jonkman J, Butterfield S, Musial W, and Scott G.
\newblock {Definition of a 5-MW Reference Wind Turbine for Offshore System
  Development}.
\newblock Technical Report NREL/TP-500-38060, National Renewable Energy
  Laboratory, Februrary 2009.

\bibitem[Kooijman et~al.(2003)Kooijman, Lindenburg, Winkelaar, and van~der
  Hooft]{Kooijman2003}
Kooijman H.~J.~T, Lindenburg C, Winkelaar D, and Hooft E.~Lvan~der.
\newblock {Aero-elastic modelling of the DOWEC 6 MW pre-design in PHATAS}.
\newblock Technical Report DOWEC-F1W2-HJK-01-046/9, September 2003.
\newblock public re-print of ECN-CX-01-135.

\end{thebibliography}

\end{document}